\pdfoutput=1

\documentclass[nofootinbib,prb,aps,showpacs,twocolumn,groupedaddress]{revtex4-1}

\usepackage{braket}
\usepackage{bbm}
\usepackage{tikz}
\usepackage{calrsfs}
\usepackage{mathtools}
\usepackage{subfigure}
\usepackage{amsmath}
\usepackage{amssymb}
\usepackage{physics}
\usepackage{hyperref}
\DeclareMathAlphabet{\pazocal}{OMS}{zplm}{m}{n}

\newcommand{\f}[2]{\frac{#1}{#2}} 
\newcommand{\beq}{\begin{equation}}
\newcommand{\eeq}{\end{equation}}

\newcommand{\bo}[1]{\boldsymbol{#1}}

\newcommand{\D}{\mathrm{d}}

\newcommand{\I}{\mathrm{I}}

\makeatletter
\renewcommand*\env@matrix[1][*\c@MaxMatrixCols c]{%
  \hskip -\arraycolsep
  \let\@ifnextchar\new@ifnextchar
  \array{#1}}
\makeatother

\newcommand{\note}[1]{\marginpar{$\color{red}\bullet$}}

\begin{document}


\title{Size Constraints on Majorana Beamsplitter Interferometer: \\ Majorana Coupling and Surface-Bulk Scattering}

\author{Henrik Schou R{\o}ising}
\email[]{henrik.roising@physics.ox.ac.uk}
\author{Steven H. Simon}
\affiliation{Rudolf Peierls Center for Theoretical Physics, Oxford OX1 3NP, United Kingdom}


\date{\today}

\begin{abstract}
Topological insulator surfaces in proximity to superconductors have been proposed as a way to produce Majorana fermions in condensed matter physics.  One of the simplest proposed experiments with such a system is Majorana interferometry. Here, we consider two possibly conflicting constraints on the size of such an interferometer. Coupling of a Majorana mode from the edge (the arms) of the interferometer to vortices in the center of the device sets a lower bound on the size of the device. On the other hand, scattering to the usually imperfectly insulating bulk sets an upper bound.  From estimates of experimental parameters, we find that typical samples may have no size window in which the Majorana interferometer can operate, implying that a new generation of more highly insulating samples must be explored. 
\end{abstract}


\pacs{71.10.Pm, 74.45.+c, 73.23.-b}


\maketitle

\section{Introduction}
There has been an ongoing search for Majorana fermions in condensed matter systems which has been intensified over several years. \cite{fermMajo12} 
Vortices in a spinless $p$-wave superconductor have long been known to bind zero energy Majorana modes.\cite{ReadGreen, Kopnin, Volovik1999, VorticesPWave} With $p$-wave superconductors being very rare in nature, no experiment has convincingly observed such Majoranas yet.\cite{Kallin}  More recently it was predicted that Majorana bound states also exist in vortices in a proximity-induced superconductor on the surface of a topological insulator (TI).\cite{FuKane_proximity}  A number of recent experiments on TIs in proximity to superconductors\cite{TopSuper, RobustFabryPerotBiSE,ProximityIndGap2ZBCP,ProximityArtificialSC,ProximityIndGapZBCP,Nature_proximityBi2Se3}  and other similar experimental systems\cite{Mourik1003, NadjPerge602} have increased the interest in this possibility.


\begin{figure}[h!tb]
\centering
\includegraphics[width=\linewidth]{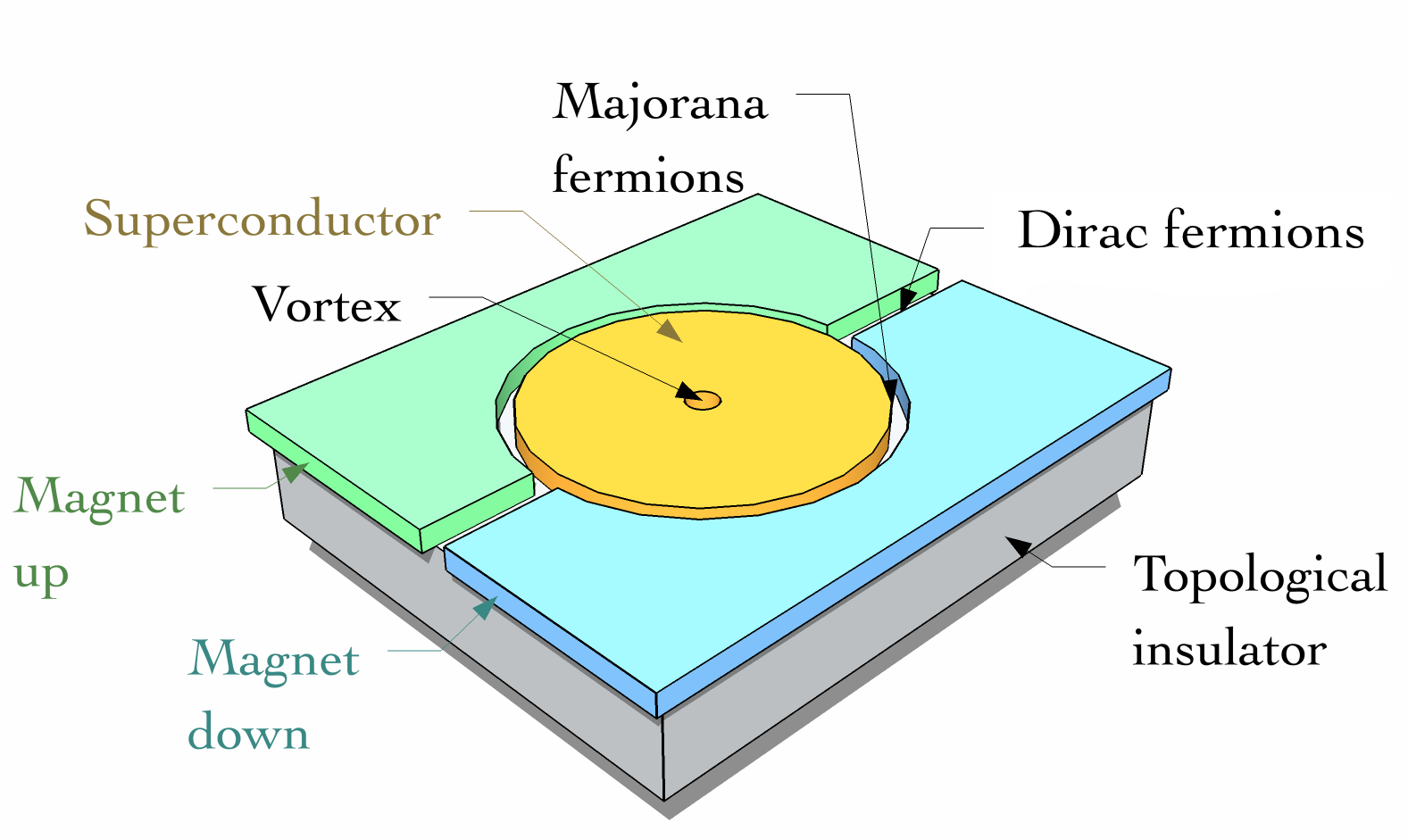}
\caption{The Majorana beamsplitter interferometer proposed by Fu and Kane, and Akhmerov et al.\cite{InterferometryBeenakker, FuKane_interferometer} It consists of a 3D strong TI in proximity to a superconductor and magnets of opposite polarization.}
\label{fig:Interferometer3D}
\end{figure}

The surfaces of TIs support gapless excitations with the dispersion of a Dirac cone\cite{TopInsulators} (in principle, TIs can have any odd number of Dirac cones in the Brillouin zone but we consider the simplest case of a single cone for simplicity). The spectrum can be  gapped either by applying a magnetic field to give the Dirac fermion either a positive or negative mass, or by placing a superconductor in proximity to the surface.  At interfaces between different gapped regions, gapless one-dimensional fermionic channels can develop.   For example, an interface between two magnetically gapped regions with opposite mass signs will contain a gapless and chiral one-dimensional Dirac fermion mode.  An interface between a magnetically gapped region and a superconducting region will contain a gapless chiral Majorana mode.\cite{FuKane_proximity,InterferometryBeenakker, FuKane_interferometer}  It is the physics of these modes that we are exploring in the current paper.

A very elegant experiment, building an interferometer out of these gapless chiral modes, was proposed by Fu and Kane\cite{FuKane_interferometer} and simultaneously by Akhmerov et al.\cite{InterferometryBeenakker}   The device is depicted schematically in Fig.\@ \ref{fig:Interferometer3D}. Incoming particles or holes, biased at a low voltage, flow into a Dirac channel between two oppositely polarized magnetic regions.   The Dirac fermion is split into two Majorana fermions upon hitting the superconductor, one flowing in each direction around the superconducting region (drawn as a disk in Fig.\@ \ref{fig:Interferometer3D}).  At the other end of the superconducting region the two Majorana modes are re-combined into a Dirac mode.    The differential conductance of this device was predicted to take the values $0$ or $2e^2/h$ depending on whether the number of $\Phi_0 = h/(2e)$ vortices in the superconductor is even or odd, respectively. With the exception of quantum Hall systems, this was the first proposed realistic Majorana interferometry experiment, and it remains a good candidate for the first experiment to successfully establish the existence of chiral Majorana modes (although, we note that promising evidence of chiral Majorana modes was reported very recently\cite{HeChiralMajorana17}).

The most experimentally explored TI materials are the Bismuth based compounds,\cite{AndoReview,TopInsFundPersp,TopInsulators,TopInsulators} including (among many others) Bi$_2$Se$_3$, Bi$_2$Te$_3$, and Bi$_2$Te$_2$Se.  Within this class of materials, several experiments have successfully formed some sort of superconducting interfaces.\cite{TopSuper, RobustFabryPerotBiSE, ProximityIndGap2ZBCP, ProximityArtificialSC, ProximityIndGapZBCP, Nature_proximityBi2Se3} In this paper we mainly have this type of material in mind.  However, we note that the material SmB$_6$,\cite{ SmB6_PRB, SmB6_PhysRevX} which is possibly a topological Kondo-insulator, will be discussed in the conclusion.

\begin{figure}[h!tb]
	\centering
	\subfigure[]{\includegraphics[width=0.46\linewidth]{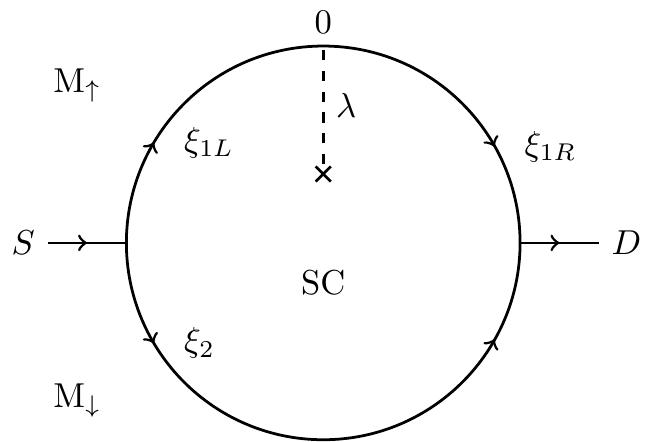} } \quad \subfigure[]{\includegraphics[width=0.46\linewidth]{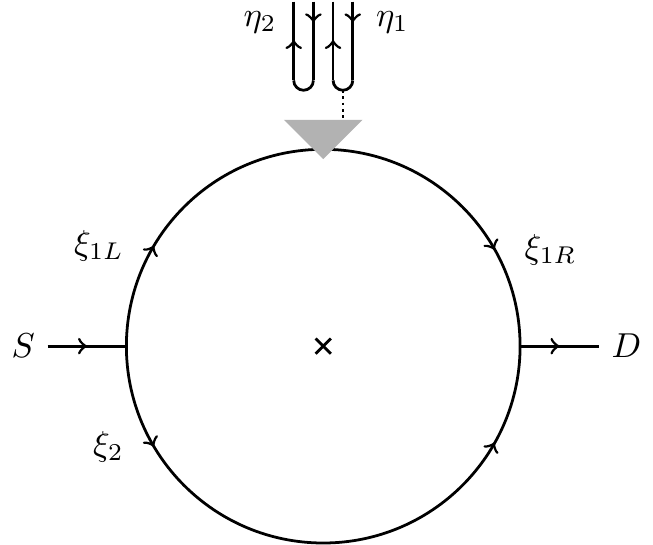}}

	\caption{Top view of the interferometry device with a superconductor (SC) and magnetic domains (with magnetization $\mathrm{M}_{\uparrow/\downarrow}$) that probes charge transport from the source (S) to the drain (D) through a pair of chiral Majorana modes. (a) Single-point coupling of strength $\lambda$ at coordinate $x=0$. Here, $\xi_{1R} \equiv \xi_1(x = 0^+)$ and $\xi_{1L} \equiv \xi_1(x = 0^-)$.
		(b) The interferometer coupled to a single-mode  conducting lead representing leakage to the bulk of the TI.}
	\label{fig:VortexEdgeCouplingSchematic}
\end{figure}

In this paper we study two effects that restrict the size of the interferometer device, as summarised in Fig.\@ \ref{fig:VortexEdgeCouplingSchematic}. On the one hand, we consider coupling of the Majorana edge mode to the Majorana modes trapped in the core of the vortex in the center of the interferometer.    When this coupling is sufficiently strong, i.e.\@ when the interferometer is sufficiently small, the conductance signal will be distorted, rendering the interpretation of the experiment difficult.   The coupling is generally strong on the scale of the coherence length $\xi=\hbar v_F/\Delta_0$ where $v_F$ is the Fermi velocity and $\Delta_0$ is the proximity gap.   This length scale can be on the order of a micron (we discuss materials parameters in section \ref{sec:LowerBoundSize}, \ref{sec:SBRate}, and \ref{sec:ContradictingRequirements}).

Next, we consider surface-bulk scattering due to the unintentionally doped and poorly insulating bulk of TIs.\cite{PRLNonmetallicBismuthSelenide, AndoReview, Skinner2013}  We find that when current leaks from the Majorana edge channel to the ground, the signal obtained at the end of the interferometer drops exponentially with a length scale set by the surface-to-bulk scattering length from disorder. For typical samples this length scale can be shorter than a micron. Thus we have two potentially conflicting constraints on the size of the proposed interferometer.   We will discuss the possible directions forward in the conclusion.

The outline of this paper is as follows. In section \ref{sec:Background} we review the formalism of Majorana interferometry and the characteristic conductance signal of the experiment. In section \ref{sec:MajoranaCoupling} we consider the impact of coupling between the chiral Majorana and a bound Majorana mode in a vortex core. We study how this limits the source-drain voltage and sets a lower bound on the size of the system. In section \ref{sec:SBScattering} we consider the leakage of current from the device to the TI substrate and we establish an upper limit on the system size due to this leakage.   We conclude with numerical estimates and an outlook for Majorana interferometry on the surface of TIs. Appendix \ref{sec:AverageCurrentDerivation} contains a derivation of the average currents. In Appendix \ref{sec:GeneralizedMajoranaCouplings} we provide generalizations of the single-point Majorana coupling displayed in Fig.\@ \ref{fig:VortexEdgeCouplingSchematic} (a). Vortex-bound Majorana fermions in a TI/SC hybrid structure and the energy splitting between zero modes is discussed in Appendix \ref{sec:MajoranaBoundStates}. In Appendix \ref{sec:SBScatteringAppendix} we present details of our estimate of the surface-bulk scattering rate based on Fermi's Golden rule. Finally, in Appendix \ref{sec:AcousticPhonons} signal loss due to scattering with acoustic phonons is briefly discussed.

\section{Background on Interferometry with Majorana Fermions}
\label{sec:Background}
In this section, we first review the formalism needed to calculate the conductance and interferometry current.\cite{FuKane_interferometer, InterferometryBeenakker, MajoResonantAndreevReflection, ScatteringChiralMajorana}  In Fig.\@ \ref{fig:VortexEdgeCouplingSchematic} (a) we show a top view of the interferometer that was described in the introduction.   For the moment we ignore the Majorana coupled to the top arm (marked as an ``X") at position 0.    Charge transport from the source to the drain is computed by using a transfer matrix which describes transport of particles and holes from the source on the left ($L$), via the perimeter of the superconductor, to the drain on the right ($R$), $[ \psi_e, \psi_h]_R^T = \pazocal{T}[\psi_e, \psi_h]_L^T$ where ${}^T$ here means transpose.   The matrix $\pazocal{T}$ can be decomposed into three pieces corresponding to the three key steps between the source and the drain
\begin{equation}
\label{eq:TransferMatrix}
\pazocal{T} = S^{\dagger}
\pazocal{P} S =  \begin{pmatrix}
\pazocal{T}_{ee} & \pazocal{T}_{eh} \\
\pazocal{T}_{he} & \pazocal{T}_{hh}
\end{pmatrix}.
\end{equation}
The unitary matrix $S$ relates the Majorana states $[ \xi_1, \xi_2]$ running along the upper and lower edge of the superconducting disk to the electron and hole states $[\psi_e, \psi_h]$ that enter via the leads, $[ \xi_1, \xi_2]^T = S[\psi_e, \psi_h]^T$.  The matrix $\pazocal{P}$ contains plane wave phases that the low energy chiral Majorana modes accumulate as they move along the edge of the superconducting disk.  Finally, the  matrix $S^\dagger$ reassembles the two Majoranas into outgoing electron and hole states that enter the drain. 

The matrices $S$ and $\pazocal{T}$ are functions of energy.   Due to the particle-hole symmetry, we must have $S(E) = S^{\ast}(-E) \tau_x$, where $\tau_x$ is a Pauli matrix in particle-hole space. At $E = 0$ these constraints fix $S(0)$ up to an overall phase that observables do not depend on,\cite{InterferometryBeenakker}
\begin{equation}
S(0) = \f{1}{\sqrt{2}}\begin{pmatrix}
1 & 1 \\
i & - i
\end{pmatrix} \begin{pmatrix}
e^{i\alpha} & 0 \\ 0 & e^{-i\alpha}
\end{pmatrix}.
\label{eq:SzeroEnergy}
\end{equation}
We will apply $S(0)$ with $\alpha = 0$ for convenience.  As discussed in Ref.\@ \onlinecite{FuKane_interferometer} this form is exact when the system has a left-right symmetry.  Even in cases where the system breaks this symmetry, corrections are $\pazocal{O}(E^2)$ and can thus be ignored at low temperature and low voltage (see Appendix \ref{sec:SmatrixCorrections}). We study the symmetric situation where the magnetization is  $ M_0 \equiv M_{\uparrow} = -M_{\downarrow}$ throughout the main text. The magnetization enters the model Hamiltonian given in Eq.\@ \eqref{eq:Model_Hamiltonian}.

The transfer matrix is used to calculate the average current in the drain as the difference between the electron and hole current, with the source biased at voltage $V$ (see Appendix \ref{sec:AverageCurrentDerivation} for a derivation),
\begin{equation}
I_D = \f{e}{h} \int_0^{\infty} \D E \hspace{1mm} \delta f(E) \left( \lvert \pazocal{T}_{ee}\rvert^2 - \lvert \pazocal{T}_{eh}\rvert^2 \right).
\label{eq:DrainCurrent}
\end{equation}
Here, $\delta f =  f_e - f_h$ with $f_{e/h}(E) = f(E\mp eV)$, and $f(E) = (1+e^{\beta E})^{-1}$ is the Fermi-Dirac distribution with $\beta = 1/(k_B T)$ and $E$ measured relative to the Dirac point. The incoming current is
\begin{equation}
I_S = \f{e}{h} \int_0^{\infty} \D E \hspace{1mm} \delta f(E).
\label{eq:SourceCurrent}
\end{equation}
By current conservation, a net current of $I_{\mathrm{SC}} = I_S - I_D$ is absorbed by the grounded superconductor. As follows from  Eq.\@ \eqref{eq:DrainCurrent}, \eqref{eq:SourceCurrent}, and unitarity of $\pazocal{T}$ the differential conductance measured in the grounded superconductor at zero temperature is:
\begin{equation}
G_{\mathrm{SC}}(V)  = \f{\D I_{SC}}{\D V} \Big\lvert_{T = 0} = \f{2e^2}{h} \lvert \pazocal{T}_{eh}(eV) \rvert^2.
\label{eq:DifferentialConductance0}
\end{equation}
In order to calculate $G_{\mathrm{SC}}(V)$ we need only establish the properties of the propagation matrix $\pazocal{P}$.   If the arms of the interferometer are of length $l_1$ and $l_2$ we have
\begin{equation}
\pazocal{P}(E) =
\begin{pmatrix}
e^{ik(E) l_1 + 2i\phi(E)} & 0 \\
0 & e^{ik(E) l_2}
\end{pmatrix}.
\label{eq:PhaseMatrix}
\end{equation}
Above the wavevector is of the form $k(E) = E/v_m$ due to the linear dispersion of the Majorana modes\cite{FuKane_interferometer} where $v_m$ is the Majorana velocity.  We have included an additional phase shift $\phi$ which may come from a number of sources.  In Refs. \onlinecite{FuKane_interferometer, InterferometryBeenakker}  the possibility of $n$ vortices being added in the center of the superconducting region was considered.  In this case the additional phase is  $e^{2 i\phi}=(-1)^n$.   Inserting $\pazocal{P}$ into Eq.\@ \eqref{eq:TransferMatrix} and \eqref{eq:DifferentialConductance0} at zero temperature yields the conductance
 \begin{equation}
 G_{\mathrm{SC},0}(V) = \frac{2e^2}{h}\sin^2\left( \f{n\pi}{2} + \f{eV \delta L}{2\hbar v_m}  \right)  .
 \label{eq:DifferentialConductanceOnePrime}
 \end{equation}
Here, $\delta L = l_1 - l_2$ is the difference between the lengths of the two arms. At $\delta L = 0$, $G_{\mathrm{SC},0}(V=0) = 2 e^2/h$ for $n$ odd, and is zero for $n$ even.  This would be a rather clear experimental signature.  For the same phase matrix $\pazocal{P}$, the drain current in Eq.\@ \eqref{eq:DrainCurrent} evaluates to
\begin{equation}
I_{D,0}(V) = (-1)^n \pi k_B T \f{e }{h} \f{\sin{(\f{eV \delta L}{\hbar v_m} )}}{\sinh{(\f{\pi \delta L k_B T}{\hbar v_m} )}} , 
\label{eq:CurrentZerothOrder}
\end{equation}
which holds in the low temperature and low voltage limit (compared to the bulk gap). We emphasise that these results are derived with zero coupling to the central Majorana and to any other degrees of freedom (e.g.\@ phonons or conducting bulk states), hence the subscript $0$.

\section{Effect of Majorana Coupling}
\label{sec:MajoranaCoupling}
We now consider the effect of coupling the Majoranas trapped in the cores of vortices in the superconductor to the chiral edge states.  Each vortex traps a single Majorana mode.   If a vortex is close to the edge (roughly within the coherence length) there will be tunnelling coupling as shown in in Fig.\@ \ref{fig:VortexEdgeCouplingSchematic} (a)   where the Majorana zero mode is marked as an ``X'' and the (tunnelling) coupling matrix element is of magnitude $\lambda$. The magnitude of the coupling drops exponentially with the distance between the vortex and the edge, see Appendix \ref{sec:VortexBoundStates} and \ref{sec:Energysplittingpplusip}.

\begin{figure}[h!tb]
\centering
\includegraphics[width=0.8\linewidth]{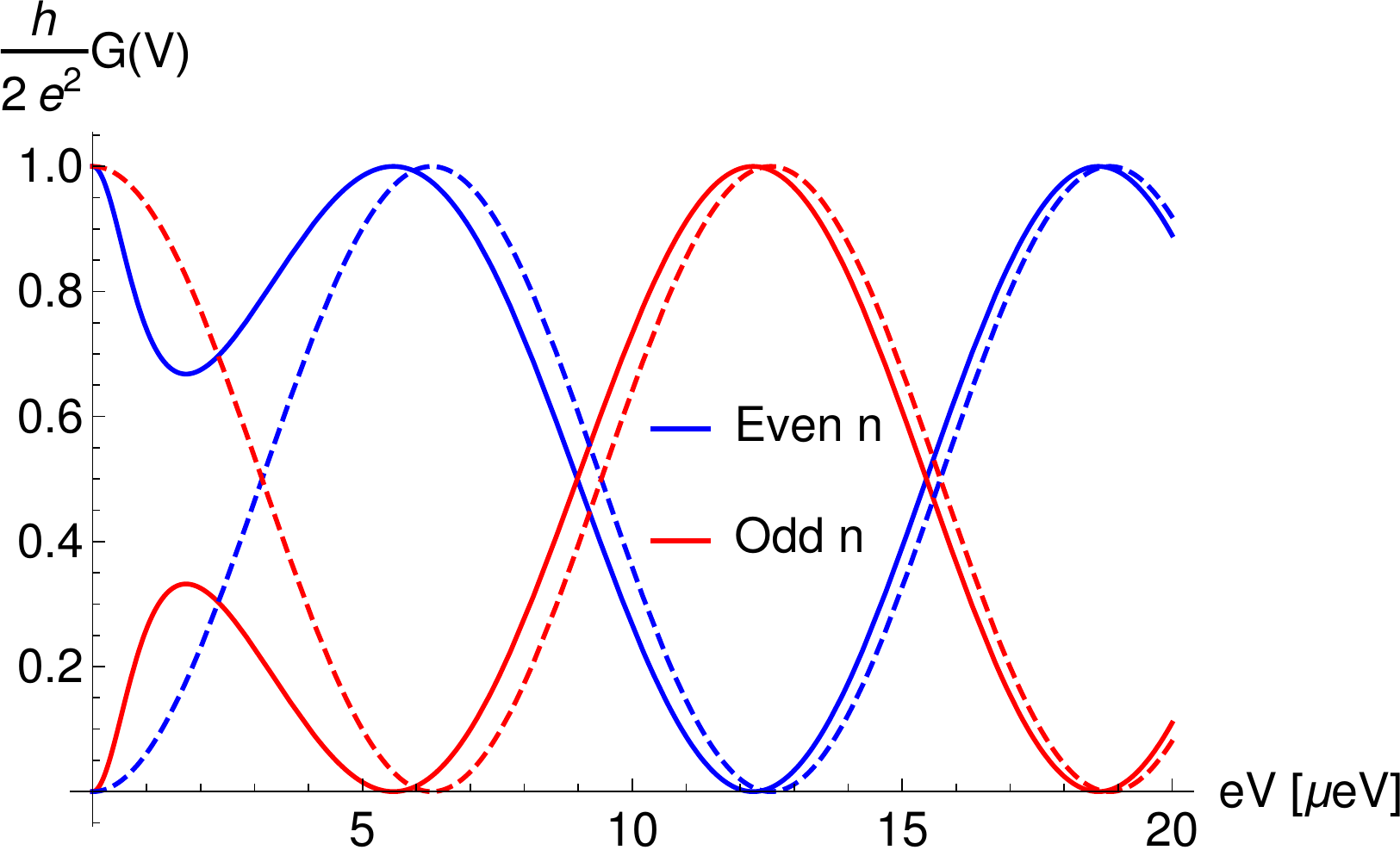} 
\caption{The differential conductance from Eq.\@ \eqref{eq:DifferentialConductanceOne} as a function of the voltage. Here, $\nu = 1$ $\mu$eV and $\hbar v_m / \delta L = 2$ $\mu$eV. The dotted lines show the conductance with $\nu = 0$. At low voltage the even-odd effect is reversed.}
\label{fig:DiffConductance}
\end{figure}

We very generally describe the chiral Majorana on the upper interferometer arm, $\xi_1$, by a Lagrangian density $\pazocal{L}_1 = i\xi_1(\partial_t + v_m \partial_x) \xi_1$ where $x$ is the spatial coordinate along the upper edge. A similar description holds for the state on the lower edge, $\xi_2$. The vortex bound Majorana, $\xi_0$, is described by $\pazocal{L}_0 =  i \xi_0 \partial_t \xi_0 $. We add the coupling term between the central bound state and the chiral mode
\begin{equation}
\pazocal{L}_{\mathrm{bulk}-\mathrm{edge}} =   2 i \lambda \xi_1(x=0) \xi_0.
\label{eq:CouplingLagrangian}
\end{equation}
The equations of motion, following from the full Lagrangian $\pazocal{L}_0 + \pazocal{L}_1 + \pazocal{L}_{\mathrm{bulk}-\mathrm{edge}}$, are given by \cite{Fendley2009, Bishara2009}
\begin{align}
2\partial_t \xi_0 &= \lambda \left[ \xi_{1R} + \xi_{1L} \right],  \\ 
v_m \xi_{1R} &= v_m \xi_{1L} + \lambda \xi_0.
\label{eq:EquationsOfMotion0}
\end{align}
Here, the notation $\xi_{1R} = \xi_1(x = 0^+)$ and $\xi_{1L} = \xi_1(x = 0^-)$ was introduced. A Fourier transformation yields a phase shift across the coupling point,
\begin{equation}
\xi_{1R}(\omega) = \f{\omega + i\nu}{\omega - i\nu} \hspace{1mm} \xi_{1L}(\omega) = e^{2i\phi} \hspace{1mm} \xi_{1L}(\omega),
\label{eq:FrequencyShift}
\end{equation}
with $\omega$ the frequency, $\nu \equiv \lambda^2/(2\hbar v_m)$, and $\phi(\omega) = \arctan(\nu/\omega)$.  This energy-dependent phase shift is inserted into Eq.\@ \eqref{eq:PhaseMatrix} and we obtain the zero-temperature result
\begin{equation}
G_{\mathrm{SC}}(V) = \f{2e^2}{h}\sin^2\left( \f{n\pi}{2} + \f{eV \delta L}{2\hbar v_m} + \arctan{\left(\f{\nu}{eV} \right) } \right).
\label{eq:DifferentialConductanceOne}
\end{equation}
Observe that the even-odd effect undergoes a crossover when the coupling strength is of the order $\lambda \approx \sqrt{2\hbar v_m eV}$. At zero voltage, or at infinite coupling strength, the even-odd effect is reversed from the value at high voltage or low coupling strength. This crossover is equivalent to shifting $n$ by one in $G_{\mathrm{SC,0}}$, causing $\xi_1$ to acquire a phase shift of $\pi$ at low energy, and it is assigned the interpretation that a vortex Majorana effectively is absorbed by the edge.\cite{Fendley2009} Similar results are known from Quantum Hall interferometers at filling fraction $5/2$. \cite{QHI52PRL2008, QHInterferometers52, Bishara2009} The original conductance is recovered at high voltage (see Fig.\@ \ref{fig:DiffConductance}). 

The above result applies when there is no position dependence in the coupling to the edge. If we instead consider a continuous bulk-edge coupling 
\begin{equation}
\pazocal{L}_{\mathrm{bulk}-\mathrm{edge}} =   2 i \int \D x \hspace{1mm} \lambda(x) \xi_1(x) \xi_0,
\label{eq:CouplingContinuous}
\end{equation}
then the total phase shift is again given by Eq.\@ \eqref{eq:FrequencyShift} with
\begin{equation}
\lambda^2 \to  2 \int \D x \hspace{1mm} \lambda( x ) e^{i k x} \int_{x} \D x' \lambda(x') e^{- i k x'}
\label{eq:CouplingReplacement}
\end{equation}
in the numerator and a similar replacement in the denominator, see  Appendix \ref{sec:SmearedCoupling}.  We note that if $x_c \ll 1/k$, where $x_c$ is defined such that $\lambda( \lvert x \rvert > x_c ) \approx 0$, then the effective $\lambda^2$ above becomes $ \langle \lambda(x) \rangle^2$ with $\langle \lambda(x) \rangle = \int \D x \lambda(x)$ up to corrections of order $kx_c$.

The scheme above can also be generalized to include more complicated couplings to multiple (vortex) Majorana modes.  So long as these modes are coupled to only a single edge, and not to each other, each coupling causes a phase shift of $\arctan(\nu/(eV))$ in the phase of propagation along the edge, see Appendix \ref{sec:CouplingMultipleVortices}.  In the case that multiple vortex Majoranas are coupled to each other, an even number of Majoranas will gap out, whereas an odd number will leave a single effective Majorana zero mode.

In the above calculation we assumed Majorana coupling to one edge only.  Since coupling varies exponentially with distance to the edge it is not unreasonable that this will effectively be the case.  However, it is also realistic that a vortex will be roughly equal distance from, and hence equally coupled to, both edges.  This case is discussed in detail in Appendix \ref{sec:CouplingBothEdges}.  While the general result becomes complicated, at least in the case of equal couplings to both edges and both edges of equal length, the physics of the even-odd crossover found in this section remains unchanged. Finally, coupling between a single vortex Majorana and the edge at finite temperature is discussed in Appendix \ref{sec:CouplingFiniteTemperature}.

\subsection{Lower Bound on Size and Voltage Constraints}
\label{sec:LowerBoundSize}
The above derived crossover makes the observation of the even-odd conductance effect impossible at low voltage.   The proposed experiment is to add a single vortex and observe a change in conductance (say, from zero to $2e^2/h$).   However, if the Majorana is then effectively absorbed into the edge, the conductance remains zero even once the vortex is added, destroying the predicted effect. This will occur for interferometers of size comparable to the coherence length, i.e.\@ the length scale of an order-parameter deformation. Assuming that $\Delta_0 = 0.1$ meV is an achievable proximity-induced gap in e.g.\@ Bi$_2$Se$_3$,\cite{ProximityIndGapZBCP, ProximityIndGap2ZBCP} the naive lower size bound for the disk is $\xi = \hbar v_F/\Delta_0 \simeq 4$ $\mu$m.\cite{NatureBismuthFermiVel} The tunnelling coupling is expected to decay like $\lambda \propto \mu \exp(- R/\xi)$ when $R\gg \xi$ (see Eq.\@ \eqref{eq:EnergySplit}) where $\mu$ is the chemical relative to the Dirac point.\cite{MajoranaSupercondIslands}  For disks of size $R \simeq \xi$ the energy splitting is comparable to the energy gap and the notion of stable edge/vortex states breaks down.

We note, however, that the Majorana coupling term vanishes identically at $\mu = 0$ in the topological insulator superconductor hybrid structure.\cite{TunnelingMajoranamodes, StronglyInteractingMajoranas} This is related to appearance of an additional symmetry at the Dirac point, bringing the Hamiltonian (Eq.\@ \eqref{eq:Model_Hamiltonian} in the absence of a magnetic field) from symmetry class D to the BDI in the Altland-Zirnbauer classification. If there is disorder inducing local fluctuations in the chemical potential (see section \ref{sec:ContradictingRequirements}),\cite{BandFluctuationsPuddlesNat} say on some scale $\delta \mu$, a random coupling term of the type in Eq.\@ \eqref{eq:CouplingContinuous} will be present and cause energy splitting. Still, the scenario of having the average $\langle \lambda(x) \rangle \sim \langle \mu(x) e^{-r(x)/\xi} \rangle \approx 0$, which would make the parasitic phase shift vanish, is possible but becomes extremely geometry sensitive (e.g. sensitive to the vortex position) if $R \simeq \xi$. Moreover, 
we should expect $\langle \lambda(x) \rangle $ to be on the scale of $\delta \mu  \sqrt{l/d}$, which in general will not be small. Here, $d$ is the length scale associated with the energy fluctuations.  Assuming that we cannot control disorder, the only way to assure suppression of  unwanted phase shifts and energy splitting is to increase $R$. Thus, we will use $R = \xi$ as a strict lower size bound on the device.

As far as chiral transport on the interferometer arms is concerned, it seems at this stage that one can operate at high voltage, $eV \gtrsim \nu$  to avoid the coupling.  Naturally, the voltage is constrained from above by the global bulk gap, $eV \lesssim \min \lbrace M_0, \Delta_0 \rbrace$, to avoid excitation of non-topological states. Thus, if $R \simeq \xi$ the remaining voltage range, $\nu \lesssim eV \lesssim \min \lbrace M_0, \Delta_0 \rbrace$, might be too limited to have a clear experimental signature. Increasing the disk radius to suppress the coupling (and therefore lower $\nu$) induces the problem of signal leakage to conducting bulk states, which we estimate below.


Finally, we note that the upper voltage bound in practice can be lower. This is due to the Caroli-de Gennes-Matricon excited states of the vortex, characterised by a minigap, $\sim \Delta_0^2/\mu$,\cite{CdGM, CdGMstatesPwaveFluids, VortexCdGMstates} which can be less than one mK. The excited vortex bound states can be activated thermally or by tunnelling from the edge states, in analogy to the zero mode tunnelling in the beginning of this section. By pinning the vortex to a hole in the superconductor, the minigap can be increased to a substantial fraction of $\Delta_0$. \cite{PinnedVortex, RobustMajoranaSCislands} Although the details of such a tunnelling process is outside the scope of this paper, additional resonances and conductance phase shifts are expected as $eV$ hits the bound state energies.  


\section{Surface-Bulk Scattering}
\label{sec:SBScattering}
Topological qubits are intrinsically protected from decoherence. \cite{AnyonQuantumComputation} In protocols based on braiding with topological qubits, leakage of current is harmful since it generically causes entanglement with the environment that potentially corrupt the qubits.\cite{FuKane_proximity} Although the experiment we study here does not probe a topological qubit, both types of experiment are sensitive to bulk leakage. Since most TIs are poorly insulating,\cite{TopInsFundPersp} bulk leakage is a relevant problem to consider.


In this section we model leakage of Majorana modes from the the surface of the TI to its poorly insulating bulk. This is done by coupling the interferometer arm first to a single metallic lead. Then, we take multiple weakly coupled metallic leads to represent a continuously leaking environment. We combine the result of this scattering process with an estimate of the surface-to-bulk scattering rate from disorder in doped TIs. Our results suggest an upper size bound that potentially coincides with the lower bound discussed in section \ref{sec:MajoranaCoupling} for many unintentionally doped TIs.


\subsection{Scattering on Conducting Leads}
Let the upper interferometer arm be coupled to a metallic lead as depicted in Fig.\@ \ref{fig:VortexEdgeCouplingSchematic} (b). Referring to the formalism of Ref.\@ \onlinecite{ScatteringChiralMajorana}, the lead fermions are transformed to a Majorana basis, $[\eta_1^{(\pm)}, \eta_2^{(\pm)}]^T = S[\psi_e^{(\pm)}, \psi_h^{(\pm)}]^T$ with the superscript indicating incoming ($-$) or outgoing ($+$) states. The $S$ matrix here may differ from the one in Eq.\@ \eqref{eq:SzeroEnergy} at low energy only by having a different phase $\alpha$, which is irrelevant for observables. The scattering process in the Majorana basis is denoted by $\pazocal{A}(E)$,
\begin{equation}
\begin{pmatrix}
\xi_{1R}, & \eta_1^{(+)}, & \eta_2^{(+)} 
\end{pmatrix}^T
= \pazocal{A}(E) \begin{pmatrix}
\xi_{1L}, & \eta_1^{(-)}, & \eta_2^{(-)} 
\end{pmatrix}^T.
\label{eq:ScatteringSingleLead}
\end{equation}
By rotating the lead particles and holes into the appropriate Majorana basis, the chiral Majorana mode can be shown to decouple from one of the (artificial) lead Majoranas.\cite{ScatteringChiralMajorana} In the low energy limit, the scattering matrix is a real rotation matrix, $ \pazocal{A}(E) \in \mathrm{SO}(3)$. If we denote the local reflection amplitude of $\eta_1$ ($\eta_2$) by $r_1$ ($r_2 = 1$), this means that the scattering matrix can be parametrized by $r_1$ at the junction only (the transmission amplitude is  $t_1 = \sqrt{1-r_1^2}$),
\begin{equation}
\pazocal{A}(E) =  \begin{pmatrix}
r_1 & - t_1  \\
t_1 & r_1
\end{pmatrix} \oplus r_2.
\label{eq:RelationWandW_M}
\end{equation}
Including the state $\xi_2$ on the lower arm and the plane wave phases acquired across the interferometer, we obtain the matrix $\pazocal{P}$ acting on $[\xi_{1L},\xi_{2},\eta_1^{(-)}, \eta_2^{(-)}]^T$,
\begin{equation}
\pazocal{P} = \begin{pmatrix}
r_1 e^{ikl_1 + in\pi} & 0 & - t_1 e^{i\f{kl_1}{2}} \\
0 & e^{ikl_2} & 0 \\
t_1 e^{i \f{kl_1}{2}+ in\pi} & 0 & r_1 \end{pmatrix} \oplus 1.
\label{eq:PmatrixScattering}
\end{equation}
Here, $r_2 = 1$ was used. The transfer matrix can be computed, and it has the $2\times 2$ sub block structure
\begin{equation}
\pazocal{T} = \big( S^{\dagger}
\oplus S^{\dagger} \big)
\pazocal{P} \big( S
\oplus S \big)
= \begin{pmatrix}[c | c]
\pazocal{T}_{S \to D} & \pazocal{T}_{\ell_{\mathrm{in}} \to D} \\ \hline
\pazocal{T}_{S \to \ell_{\mathrm{out}}} & \pazocal{T}_{\ell_{\mathrm{in}} \to \ell_{\mathrm{out}}}
\end{pmatrix}.
\label{eq:SingleLeadScatteringT}
\end{equation}
Above, the subscripts indicate the result of transport from/to the source ($S$), the drain ($D$), or the incoming/outgoing lead ($\ell_{\mathrm{in/out}}$). The blocks in this transfer matrix are used to find the average current contribution as measured in contact $\beta$, with
\begin{equation}
\begin{split}
&\lvert (\pazocal{T}_{\alpha \to \beta})_{ee} \rvert^2 - \lvert (\pazocal{T}_{\alpha \to \beta})_{eh} \rvert^2 \\ 
 &= \begin{cases}
r_1(-1)^n\cos{\left( \f{E\delta L}{v_m}\right)} & (\alpha, \beta) = (S, D) \\
r_1 &  (\alpha, \beta) = (\ell_{\mathrm{in}}, \ell_{\mathrm{out}}) \\
0 & (\alpha, \beta) = (S, \ell_{\mathrm{out}}), (\ell_{\mathrm{in}}, D)
\end{cases}.
\end{split}
\label{eq:SingleLeadScatteringCurrentKernel}
\end{equation}
Here, the first line gives the contribution from combining Majoranas $\xi_1$ and $\xi_2$, the second line from combining $\eta_1$ and $\eta_2$, and the third line from combining $\xi_1$ and $\eta_1$. Combining Majoranas from different sources always give a vanishing contribution to the average current.\cite{ScatteringChiralMajorana}

The drain current is thus $I_{D}(V) = r_1 I_{D,0}(V)$, where $I_{D,0}$ is defined in Eq.\@ \eqref{eq:CurrentZerothOrder}; the visibility is coherently reduced by $r_1$. The current in the outgoing lead is $I_{\ell_{\mathrm{out}}} =  r_1 e^2 V'/h$ at $T= 0$ when the lead is biased at $V'$. Subtracting the incoming current $I_{\ell_{\mathrm{in}}} = e^2V'/h$ yields a net current of $I_{\ell_{\mathrm{out}}} - I_{\ell_{\mathrm{in}}} = (r_1 - 1) e^2 V'/h$ in the conducting lead. When decoupling the lead completely, $r_1 = 1$, no net current goes in the lead.

The matrix $\pazocal{P}$ can be trivially extended to include scattering on many single-mode leads. Repeating the calculation above with two scattering leads of the same reflection amplitude $r_1$ gives the same reduction of current in both leads, independent of which arms the leads are coupled to. The drain current is reduced by $r_1^2$.  Generalising these statements by induction,\footnote{This can be proven formally; the proof is a straightforward generalization of Eq.\@ \eqref{eq:PmatrixScattering} and \eqref{eq:SingleLeadScatteringT} when identifying the structure of the sub blocks in $\pazocal{T}$. Blocks characterizing transport between different leads always give a vanishing contribution to the average current, related to the observation right below Eq.\@ \eqref{eq:SingleLeadScatteringCurrentKernel}. If two leads are not causally connected by the chiral arm, the corresponding block in the $\pazocal{T}$ matrix is zero. Finally, no matter which arms the $N$ leads are coupled to, the drain current is reduced by the factor $\prod_{j=1}^N r_j$, where $r_j$ is the local reflection amplitude of lead $j$. } with $N$ identical scatterers of reflection amplitude $r_1$, the total conductance from the collection of leads (representing the bulk of the TI) is $G_{\mathrm{leads}} = e^2 N(1-r_1)/h$. Furthermore, the $N$ scatterers reduce the visibility of the drain current multiplicatively as $I_{D}(V) = r_1^N I_{D, 0}(V)$.

If we let $l_S$ denote the average length a chiral Majorana travels before  it scatters into the bulk, we may by definition express the reflection amplitude as $r_1 = 1 - \f{1}{N} \f{ l }{ l_S }$, which is equivalent to defining the leakage conductance into the collection of leads by $G_{\mathrm{leads}} = e^2 l/(h l_S)$. Taking the continuum limit of infinitely many weak scatterers means that $\lim_{ N\to \infty} r_1^N = \exp(- l/l_S)$, and the current is exponentially suppressed in the drain, 
\begin{equation}
I_{D}(V) =  I_{D,0}(V) e^{ - l/l_S }.
\label{eq:ReducedDrainCurrent}
\end{equation}
As for the differential conductance measured in the grounded superconductor, the amplitude of the oscillations are suppressed
\begin{equation}
G_{\mathrm{SC}}(V) = G_{\mathrm{SC},0}(V) e^{ - l/l_S  } + \f{e^2}{h} (1- e^{ - l/l_S  } ),
\label{eq:ReducedDiffConductance}
\end{equation}
and distinguishing even from odd $n$ is rendered difficult as $l$ surpasses $l_S$. Thus, $l_S$ acts as an upper bound on the circumference of the length of the interferometer arms.

\subsection{The Surface-Bulk Scattering Rate}
\label{sec:SBRate}
The remaining step of our argument is to estimate the scattering length $l_S$. We start by calculating the surface-bulk (SB) scattering length for a TI surface electron ($l_S^{(e)}$) in the absence of any surface superconductor or magnet. Then, we use a similar calculation to estimate the Majorana scattering length ($l_S^{(m)}$).  We restrict ourselves to elastic (zero temperature) surface-bulk scattering, where the electron-phonon coupling is irrelevant (see subsection \ref{sec:OtherLimitations} and Appendix \ref{sec:AcousticPhonons}). Building on the formalism developed in Ref.\@ \onlinecite{SBcouopling14} we consider scattering via screened charge impurities. We note that the consideration in Ref.\@ \onlinecite{SBcouopling14} is restricted to point scatterers where the overall scattering strength is a priori unknown, whereas in our approach the effective potential strength is fixed by the screened Coulomb potential and the dopant concentration.

\begin{figure}[h!tb]
	\centering
\subfigure[]{\includegraphics[width=0.70\linewidth]{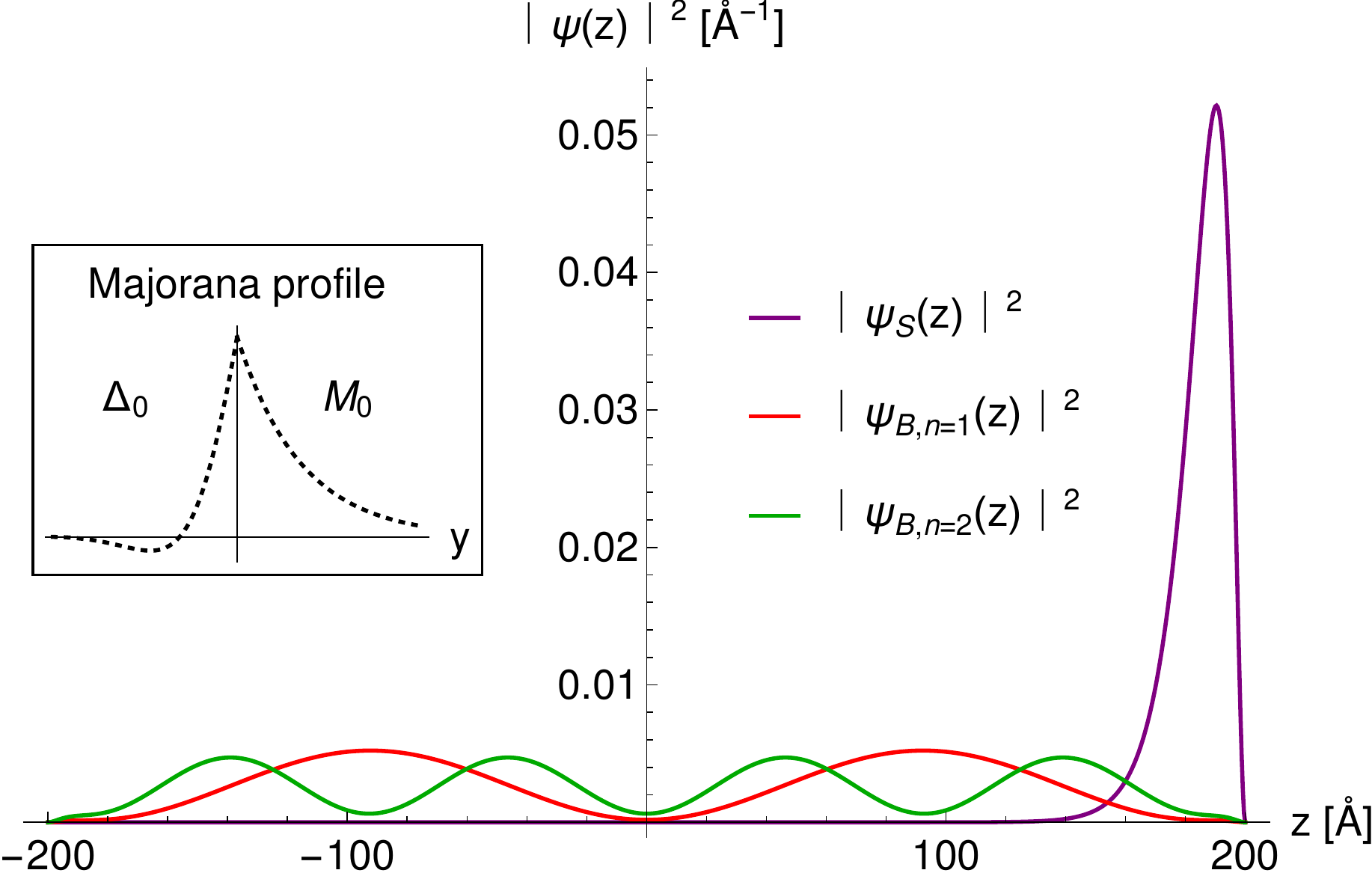}}  \quad \subfigure[]{\includegraphics[width=0.50\linewidth]{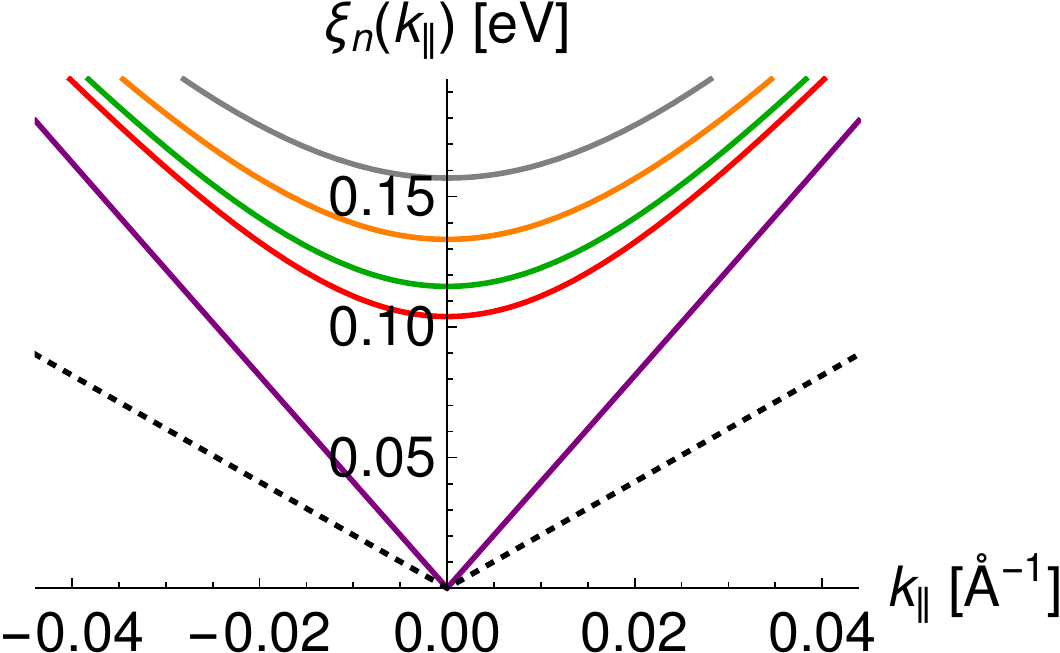} } \quad
\subfigure[]{\includegraphics[width=0.60\linewidth]{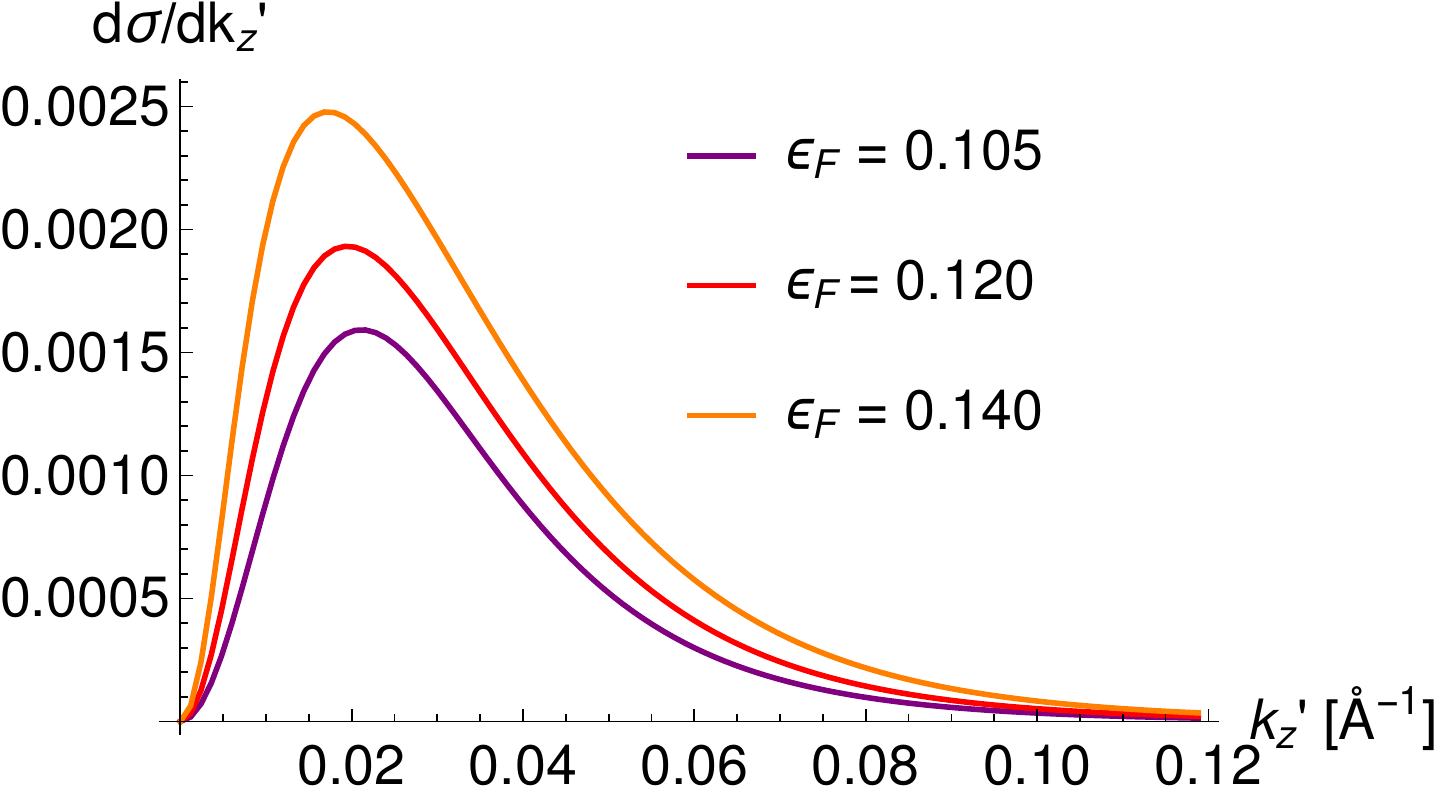} } \quad
\subfigure[]{\includegraphics[width=0.70\linewidth]{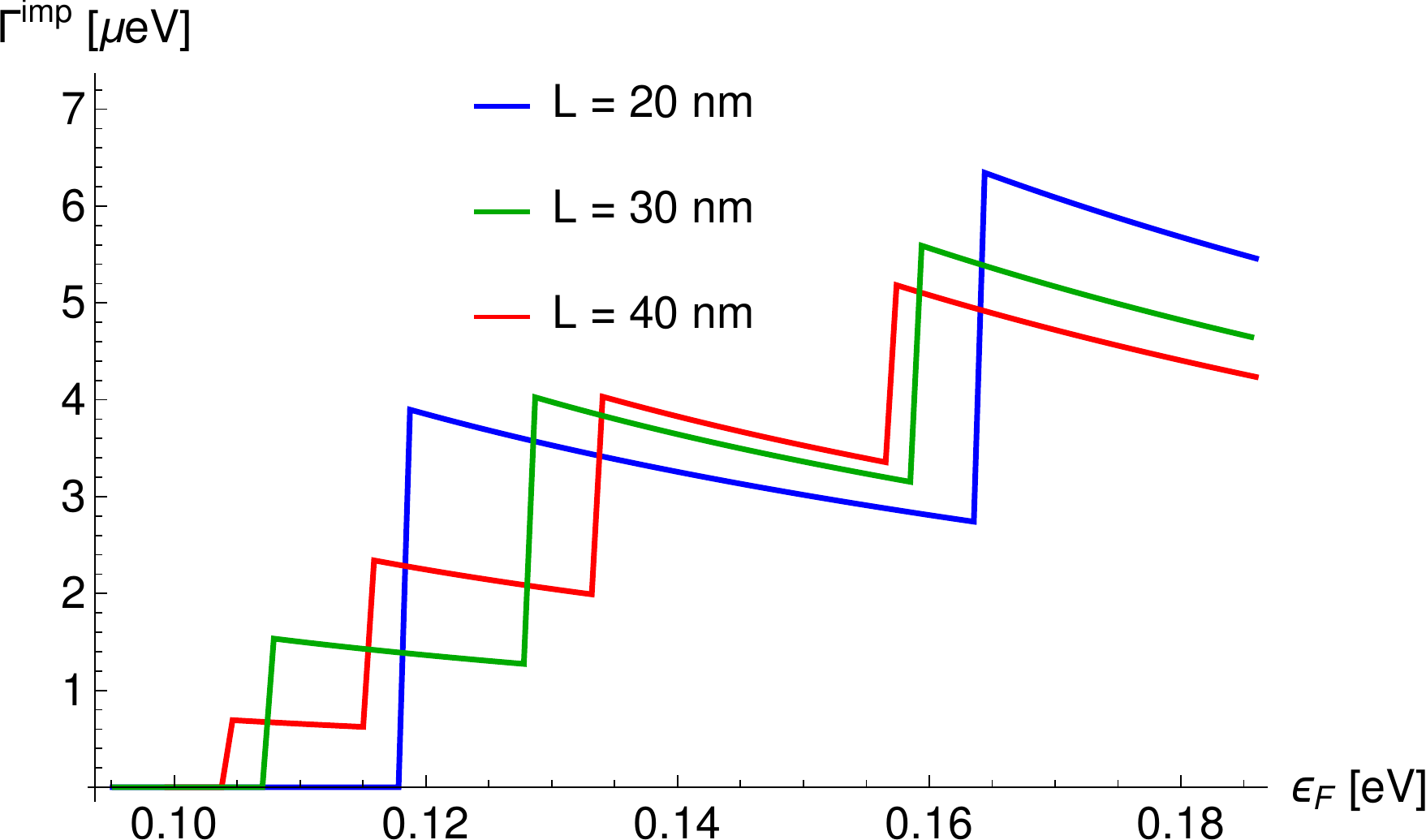}}

\caption{ (a) The three lowest-lying eigenstates in a TI film of thickness $L = 40$ nm. The surface state localized on the TI top is displayed in purple; the bottom surface state is not shown. Inset: A sketch of the typical transverse surface Majorana wavefunction. Its effect on the scattering rate is discussed in Appendix \ref{sec:ConfinedWF}. The model parameters are adjusted to those of Bi$_2$Se$_3$ except for the gap, which is here set to $\Delta_{\mathrm{TI}} = 0.1$ eV to model puddles from the conduction band.\cite{SBcouopling14, HamiltonianSurface10} (b) The dispersion of the states displayed in (a) but with two additional bulk states. In black: The typical Majorana surface dispersion. (c) The differential scattering rate $\D \sigma_{\mathrm{long}}^{(n'=1)}(\epsilon_F) / \D k_z' $ defined in Eq.\@ \eqref{eq:DifferentialCrossectionLong} shown for three values of the Fermi energy (measured in eV). (d) The electronic scattering rate as given in Eq.\@ \eqref{eq:ScatteringRateLong} with $n_{\mathrm{3D}} = 10^{19}$ cm$^{-3}$. Van Hove singularities are seen whenever the Fermi energy hits the bulk bands, but they are softened by the $k_{\parallel}'$ factor that goes to zero at these points. Here, we used $\epsilon_r = 100$.\cite{SurfImpScat10} }
\label{fig:WavefuncsAndRate}
\end{figure}

Specifically, we consider a surface state ($S$) with initial wave vector $\bo{k} = (\bo{k}_{\parallel}, 0)$ that scatters to a lowered conduction band (a charge puddle) $n'$ in the bulk ($B$) with final wave vector $\bo{k}' = (\bo{k}_{\parallel}', k_z')$. The incoming surface state has energy $\epsilon_F = v_F k_{\parallel}$. For numerical purposes we work with a TI slab of thickness $L = 40$ nm.  In Fig. \ref{fig:WavefuncsAndRate} (a) we show the form of $\lvert \Psi(z) \rvert^2$ at zero parallel momentum for the lowest three eigenstates (derived in Ref.\@ \onlinecite{HamiltonianSurface10}). One of the TI surface states, penetrating tens of \AA s into the bulk, is seen in purple. Assuming randomly distributed screened (dopant) charges with average concentration $n_{\mathrm{3D}}$ we get the scattering rate from Fermi's Golden rule (cf.\@ Appendix \ref{sec:SBScatteringAppendix}):
\begin{align}
\Gamma^{\mathrm{imp}}_{\mathrm{SB}}(\epsilon_F) &= \label{eq:ScatteringRateLong} \\
 & \hspace{-40pt}  n_{\mathrm{3D}}  \left( \f{e^2 }{2\pi \epsilon_0 \epsilon_r k_{\parallel}^2}  \right)^2   \sum_{n' \in B} \f{ k_{\parallel}' }{\lvert \nabla  \xi_{B,n',k_{\parallel}' }  \rvert} \int_0^{\infty} \D k_z' \f{\D\sigma_{\mathrm{long}}^{(n')}(\epsilon_F)}{\D k_z'} \nonumber.
\end{align}
Here, $\xi_{B,n',k_{\parallel}'}$ is the dispersion of bulk band $n'$ defining the outgoing wave vector $k_{\parallel}'$ by $\xi_{B,n',k_{\parallel}'} = \epsilon_F$. The differential scattering rate $\D \sigma_{\mathrm{long}}^{(n')}(\epsilon_F)/ \D k_z'$ is the screened Coulomb coupling convoluted with the overlap between the bulk and surface states (cf.\@ Eq\@ \eqref{eq:DifferentialCrossectionLong}), see Fig.\@ \ref{fig:WavefuncsAndRate} (c).

In most topological insulators the Fermi level resides in the bottom of the conduction band or the top of the valence band, and a significant doping of acceptors/donors is needed to reach a bulk insulating state.\cite{Nature09, Skinner2013} For Bi$_2$Se$_3$ films (typically being $n$-doped) we use the dopant concentration $n_{\mathrm{3D}} = 10^{19}$ cm$^{-3}$. \cite{GapsHybridStruct17, Skinner2013, ChargePuddlesPRB16} Such doping causes fluctuations in the average (doping) concentration, making the conduction band bend and induce electron and hole puddles,\cite{Skinner2013, BandFluctuationsPuddlesNat} i.e.\@ effectively filled pockets from the conduction band at the Fermi level. As a simple model of a typical situation we imagine that chemical doping of the TI bulk lowers the bulk Fermi level to lie in some region $0.10$  eV $ \lesssim \epsilon_F \lesssim 0.18$ eV relative to the Dirac point. This would naively be bulk insulating since the Dirac point in Bi$_2$Se$_3$ is separated from the conduction band by a gap of $0.28$ eV.\cite{NatureBi2Se3} Puddles coming from stretching the bottom of the conduction band are simply modelled by setting the TI gap to $\Delta_{\mathrm{TI}} = 0.1$ eV, making elastic scattering to the puddles (i.e.\@ the lowered conduction bands) allowed at the Fermi level, see Fig.\@ \ref{fig:WavefuncsAndRate} (b). This captures the generic features in unintentionally doped TIs.

In order to consider scattering from the surface state into the bulk, we need to know how close the surface Fermi energy is to the Dirac point. Note that, as we will see below, the scattering rate increases strongly if we are not very close (compared to $\Delta_0$). Let us therefore assume that the surface Fermi energy is tuned near the Dirac point. The electronic scattering rates as a function of the bulk Fermi energy are shown in Fig.\@ \ref{fig:WavefuncsAndRate} (d). We note that the rate stays similar in magnitude for a large TI film thickness range. When the film is made thinner, there are fewer bulk states to scatter into, but this is compensated by an increased surface-bulk overlap. Moreover, the rates in Fig.\@ \ref{fig:WavefuncsAndRate} (d) have the same qualitative features as seen for point source scattering.\cite{SBcouopling14}

The (elastic) scattering lifetime of the surface electrons is $\tau_{\mathrm{SB}}^{(e)} = 1/\Gamma^{\mathrm{imp}}_{\mathrm{SB}}$, and the scattering length is given by $l_S^{(e)} = \tau_{\mathrm{SB}}^{(e)} v_{F}$. As a typical value of the rates in Fig.\@ \ref{fig:WavefuncsAndRate} (d) we use $\Gamma^{\mathrm{imp}}_{\mathrm{SB}} = 3$ $\mu$eV, and we arrive at the unusually long scattering time $\tau_{\mathrm{SB}}^{(e)} \approx 0.2$ ns ($l_S^{(e)} \approx 0.1$ mm). This is typically a factor of $10^2-10^3$ longer than for bulk scattering lifetimes seen in experiment.\cite{ChargePuddlesPRB16} Yet, our results could be consistent with what was attributed to be unusually long surface lifetimes as observed after optically exciting bulk states.\cite{PRLElectronlifetimeBiSe, PRLElectroniclifetime2} In Ref.\@ \onlinecite{PRLElectronlifetimeBiSe} one deduced that $\tau_{\mathrm{BS}}^{(e)}>10$ ps after seeing a stable population of the surface state induced by elastic bulk-surface scattering in Bi$_2$Se$_3$ (the samples were kept at $T = 70$ K) after subsequent inelastic decays associated with much shorter time scales. It would be desirable to see similar experiments engineered to measure the elastic surface-bulk scattering rate in doped compounds conducted at lower temperatures. 


\subsection{Velocity Suppression and Contradicting Requirements}
\label{sec:ContradictingRequirements}
We now use the same expression in Eq.\@ \eqref{eq:ScatteringRateLong} to extract information about the Majorana lifetime and scattering rate.  In doing so we ignore the (confined) transverse profile of the Majorana surface state.  This is a valid approximation because the reciprocal coherence length is far less than the maximal transverse scattering momentum in the studied energy regime, see Appendix \ref{sec:ConfinedWF}.  Moreover, we assume that the induced superconductivity does not gap the TI bulk such that elastic surface-bulk scattering is not precluded. The Majorana scattering rate will typically be larger or equal to the electronic scattering rate due to a suppression of the Majorana velocity relative to the Fermi velocity. 

The Majorana velocity is calculated from the low-energy chiral solution of the Hamiltonian in Eq.\@ \eqref{eq:Model_Hamiltonian}: $v_m/v_F =  \sqrt{1-(\mu/M_0)^2} [1+(\mu/\Delta_0)^{2} ]^{-1}$.\cite{FuKane_interferometer} Here, $\mu$ is the surface chemical potential relative to the Dirac point. From the $k_{\parallel} \propto 1/v_F$ dependency of $\Gamma^{\mathrm{imp}}_{\mathrm{SB}}$ in Eq.\@ \eqref{eq:ScatteringRateLong} we deduce that $l_S^{(m)} = \tau_{\mathrm{SB}}^{(m)} v_m $ is suppressed approximatively as $l_S^{(m)} \sim (v_m/v_F)^{\alpha} \hspace{1mm} l_S^{(e)}$ with $\alpha \approx 4$ (up to some prefactor of order $1$).\footnote{The prefactor of $\Gamma^{\mathrm{imp}}_{\mathrm{SB}}$ in Eq.\@ \eqref{eq:ScatteringRateLong} has a $k_{\parallel}^{-4}$ dependency. There is, however, also a less significant dependency on $k_{\parallel}$ in $\D \sigma_{\mathrm{long}}^{(n')}(\epsilon_F) / \D k_z' $, which can be inferred from Fig.\@ \ref{fig:WavefuncsAndRate} (b) (recall that $\epsilon_F \propto k_{\parallel}$) to be very close to linear. } Correspondingly, $\tau_{\mathrm{SB}}^{(m)} \sim (v_m/v_F)^{\alpha-1} \tau_{\mathrm{SB}}^{(e)}$. A non-zero size window for the interferometer size is required by imposing $l_S^{(m)} \gtrsim \xi$, which in turn means that
\begin{equation}
v_m/v_F \gtrsim \left( \xi / l_S^{(e)} \right)^{1/4}.
\label{eq:LengthCriterion}
\end{equation}
Inserting the electronic rate from the end of the last subsection and $\xi \simeq 4$ $\mu$m in this yields $v_m/v_F \gtrsim 0.4$. Hence, we must require a very fine tuning of $\mu$, $\lvert \mu \rvert \lesssim 1.2 \Delta_0 \simeq 0.1$ meV given that one can achieve $M_0 \gg \mu$ .

When the TI bulk is increasingly doped with screened Coulomb charges to make the Fermi energy approach the Dirac point, fluctuations in the surface electrical potential energy are enlarged. The spatial dependence of the local density of states broadens the Fermi energy into a Gaussian distribution of finite width. The standard deviation of this energy smearing has been estimated in theory\cite{Skinner2013} and measured in experiment\cite{BandFluctuationsPuddlesNat} to be $\delta \mu \simeq 10 - 20$ meV within some $0.1$ eV from the average Dirac point ($\delta \mu$ decays inversely proportional to the average chemical potential further away). For Fermi energies close to the average Dirac point the notion of a uniform local density of states breaks down. Tuning of the surface chemical potential is consequently not globally possible within an energy resolution set by the distribution width.\cite{ChargePuddlesPRB16} Importantly, the smearing greatly exceeds the tuning required above, $\lvert \mu \rvert \sim \delta \mu \gg \Delta_0$, even if we relax the assumption of a very small $\Delta_0$ by increasing it one order of magnitude.

Finally, the spatial scale of the chemical potential fluctuations is $n_{\mathrm{3D}}^{-1/3} \approx 5$ nm. Hence, a precise tuning of the chemical potential might be possible within local regions of this size. Still, this would not help in probing the experiment since $n_{\mathrm{3D}}^{-1/3} \ll \xi$. Thus, unless these spatial fluctuations in the local density of states can be brought under control, unintentionally doped TIs are left unsuited for Majorana interferometry.

\subsection{Other Limitations}
\label{sec:OtherLimitations}
Majorana interactions\cite{StronglyInteractingMajoranas} and coupling between Majorana modes and other degrees of freedom, such as phonons, are other potential sources of decoherence. Local interactions between chiral Majorana fermions are expected to be heavily suppressed at low temperatures and momenta with the leading order term going like $\pazocal{O}(k^6)$. \cite{FuKane_interferometer} The chiral Majoranas on the interferometer arms can also excite phonons if an electron-phonon coupling is present; see Appendix \ref{sec:AcousticPhonons} for a short discussion. For spatial inversion symmetric materials, e.g.\@ Bi$_2$Se$_3$ and Bi$_2$Te$_3$, where this coupling is dictated by a deformation potential, the decay rate of quasiparticles due to scattering on acoustic phonons at $T=0$ exhibits a $\Gamma^{\mathrm{ph}} \sim (eV)^3$ behaviour. This is a posteriori expected to be small compared to bulk leakage at low voltage. Without spatial inversion symmetry, a piezoelectric interaction can cause the electron-phonon coupling to follow a reciprocal power law in $q$, making scattering an increasing problem for small momenta.

\section{Conclusions}
Two limiting effects in Majorana interferometers are considered. We include a previously neglected coupling between chiral Majoranas and vortex-pinned modes. At low voltage this coupling yields a crossover in the conductance even-odd effect, distorting the interpretation of the experiment. We also find that surface-to-bulk scattering in the TI sets an upper bound on the size of the device. With a proximity gap of $\Delta_0 = 0.1$ meV, the lower bound (the coherence length) is about $\xi = \hbar v_F/\Delta_0 \simeq 4$ $\mu$m for Bi$_2$Se$_3$.  Due to the doping needed to reach a bulk insulating state in many TIs, conduction band puddles lead to large fluctuations in the surface Fermi energy. In turn, this is in conflict with the surface chemical potential fine tuning required to have a non-zero size window for the interferometer. This leaves the possibility of probing the experiment in many poorly bulk insulating Bismuth compounds potentially extremely difficult.

The most natural ways to overcome the restrictions considered here are (i) to find superconductors with excellent contact to the TI such that $\Delta_0$ can be made (ideally orders of magnitude) larger, or (ii) to pursue a search for TIs with a highly insulating bulk. Most Bismuth based TIs are poorly insulating and are unintentionally doped,\cite{AndoReview} which results in them being unsuited for Majorana interferometry. However, recently reported mixed Bismuth compounds have shown evidence of an appreciably insulating bulk,\cite{ExcellentBulkResistivity} which could potentially open a window of opportunity for this material. We note that the recommendation (i) above is similar to the need for high-quality interfaces in superconductor-semiconductor nanowire devices.\cite{NanoWireReview}


Another possible material system to consider is the putative topological Kondo insulator SmB$_6$, which gives strong sign of surface conduction.\cite{SmB6_PhysRevX, SmB6_SciRep} The bulk resistivity of this material can reach several $\Omega$cm at temperatures below a few Kelvins.\cite{SmB6_PhysRevX} Very recently, evidence of the superconducting proximity effect in Nb/SmB$_6$ bilayers was reported,\cite{SmB6_ProximityEffect} hence providing a possible platform for this experiment. However, the physics of highly interacting topological Kondo insulators is still poorly understood, \cite{SmB6_Nature, SmB6_PRB} and it is unclear how much of the details of the simple non-interacting TI surface physics will carry through to this more complicated case.



\begin{acknowledgments}
We thank Mats Horsdal, Dmitry Kovrizhin, Ramil Akzyanov for useful discussions and Kush Saha for feedback on the numerics. H.\@ S.\@ R.\@ acknowledges Aker Scholarship. S.\@ H.\@ S.\@ is supported by EPSRC grant numbers EP/I031014/1 and EP/N01930X/1.
\end{acknowledgments}

\begin{appendix}

\section{Derivation of the Average Current}
\label{sec:AverageCurrentDerivation}
The derivation of Eq.\@ \eqref{eq:DrainCurrent} and \eqref{eq:SourceCurrent} goes along the lines of Ref.\@ \onlinecite{Buttiker92}. The scattering states of the interferometer are (chiral) plane waves in the one-dimensional channel convoluted with a transverse wavefunction (we let $k>0$ by definition below),
\begin{align}
\psi_L^e &= e^{ik_L x} \varphi_L^e(y), \label{eq:Scatter1} \\
\psi_L^h &= e^{-ik_L x} \varphi_L^h(y), \label{eq:Scatter2} \\
\psi_R^e &= \pazocal{T}_{ee} e^{ik_R x} \varphi_R^e(y) + \pazocal{T}_{eh} e^{-ik_R x} \varphi_R^h(y), \label{eq:Scatter3} \\
\psi_R^h &= \pazocal{T}_{hh} e^{-ik_R x} \varphi_R^h(y) + \pazocal{T}_{he} e^{ik_R x} \varphi_R^e(y). \label{eq:Scatter4}
\end{align}
We have assumed that the contacts $\alpha \in \lbrace L, R \rbrace$ contain only single mode states and that the incident state in the $L$ contact is either an electron or a hole.
\begin{figure}[h!tb]
\centering
\includegraphics[width=0.9\linewidth]{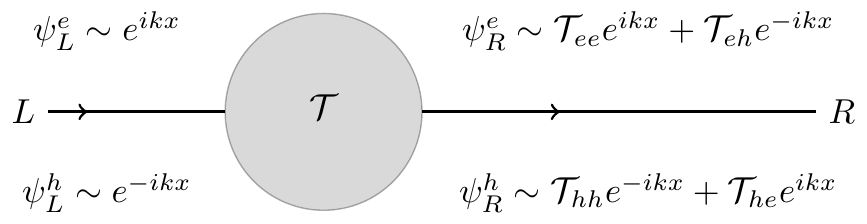}
\caption{The scattering states at the two contact points. }
\label{fig:Blackbox}
\end{figure}
One can construct arbitrary states by expanding in the scattering states above. This is incorporated in the (second quantized) field operator
\begin{equation}
\hat{\Psi}_{\alpha,\sigma}(\bo{r},t) = \f{1}{\sqrt{2\pi}} \int \f{\D E_{\alpha}}{\sqrt{\hbar v_{\alpha}}} \hspace{1mm} \psi_{\alpha}^{\sigma}(E_{\alpha},\bo{r}) \hat{a}_{\alpha,\sigma}(E_{\alpha}) e^{-i\omega_{\alpha} t},
\label{eq:FieldOperator}
\end{equation}
where we introduced $\omega_{\alpha} = E_{\alpha}/\hbar = v_{\alpha} k_{\alpha}$ and the annihilation operator $\hat{a}_{\alpha,\sigma}(E_{\alpha})$ of type $\sigma \in \lbrace e, h \rbrace$ satisfying
\begin{equation}
\big\lbrace \hat{a}_{\alpha,\sigma}^{\dagger}(E),\hat{a}_{\beta,\sigma'}(E') \big\rbrace = \delta_{\alpha,\beta} \delta_{\sigma, \sigma'} \delta(E-E').
\label{eq:CommutationsAs}
\end{equation}
The current operator of type $\sigma$ in contact $\alpha$ is defined by
\begin{equation}
\hat{I}_{\alpha,\sigma} = \f{\hbar e}{2im} \int \D \bo{r}_{\perp, \alpha} \hspace{1mm} \left[ \hat{\Psi}_{\alpha,\sigma}^{\dagger} \partial_x \hat{\Psi}_{\alpha,\sigma} - (\partial_x \hat{\Psi}_{\alpha,\sigma}^{\dagger}) \hat{\Psi}_{\alpha,\sigma} \right].
\label{eq:CurrentOperator}
\end{equation}
Here, $m$ is the effective mass, $m v_{\alpha} = \hbar k_{\alpha}$. Assuming further that the contacts act as thermal reservoirs kept at equal temperature, we can average the density operator by
\begin{equation}
\big\langle \hat{a}_{\alpha,\sigma}^{\dagger}(E)  \hat{a}_{\alpha',\sigma'}(E') \big\rangle = \delta_{\alpha,\alpha'} \delta_{\sigma,\sigma'} \delta(E - E') f_{\sigma}(E),
\label{eq:ThermalAverage}
\end{equation}
where $f_{e/h}(E) = [\exp(\beta [E \mp \mu])+1]^{-1}$ is the Fermi function for particles and holes. Finally, we assume the transverse wavefunctions to be orthonormal,
\begin{equation}
\int \D y \hspace{1mm} (\varphi_{\alpha}^{\sigma}(y))^{\ast} \varphi_{\alpha'}^{\sigma'}(y) = \delta_{\alpha,\alpha'} \delta_{\sigma,\sigma'}.
\label{eq:OrthonormalWFs}
\end{equation}
The source and the drain current are defined by the average particle minus hole current, $I_D \equiv \big< \hat{I}_{D,e} - \hat{I}_{D,h} \big>$ and $I_S \equiv \big< \hat{I}_{S,e} - \hat{I}_{S,h} \big>$. Calculating these explicitly by using Eq.\@ \eqref{eq:CurrentOperator}, \eqref{eq:ThermalAverage}, \eqref{eq:OrthonormalWFs}, and unitarity of $\pazocal{T}$ leads exactly to the expressions in Eq.\@ \eqref{eq:DrainCurrent} and  \eqref{eq:SourceCurrent}.

\section{Charge Transport with Majorana Coupling Disorder}
\label{sec:GeneralizedMajoranaCouplings}

\subsection{One Majorana with Smeared Coupling to the Edge}
\label{sec:SmearedCoupling}
Consider the case where the edge Majorana is coupled continuously to the bound state as in Eq.\@ \eqref{eq:CouplingContinuous}.  Using the ansatz $\xi_1 (x) = f(x) e^{i (k x - \omega t) } $ with $f(-\infty) $ being the Majorana (fermion) field on the far left and  $\omega = k/v_m$ for the linearly dispersing modes.  The equations of motion can be combined to give the continuum version of Eq.\@ \eqref{eq:FrequencyShift}:
\begin{equation}
- iv_m \omega \partial_x f(x) = \lambda(x) e^{- i k x} \int \D x' \hspace{1mm} \lambda(x') f(x') e^{ i k x'}.
\label{eq:SmearedCoupling}
\end{equation}
Here, the plane wave phase contribution that can be neglected in the discrete case where $\lambda(x) = \lambda \delta(x)$ is included.  The equation above can be solved as follows. Integrating both sides gives the solution implicitly by
\begin{equation}
f(x) = f(-\infty) - \f{\zeta}{i\omega v_m} \int^{x} \D x' \hspace{1mm} \lambda(x') e^{-i k x'},
\label{eq:Functionf}
\end{equation}
with $\zeta = \int \D x \lambda(x) f(x) e^{ i k x }$. Inserting this back into \eqref{eq:SmearedCoupling} yields
\begin{equation}
\zeta  =  \f{ f( -\infty)  \int \D x \hspace{1mm} \lambda(x) e^{ i k x }  }{ 1+ \f{1}{ i\omega v_m } \int \D x \hspace{1mm} \lambda( x ) e^{i k x} \int^{x} \D x' \lambda(x') e^{- i k x'}   },
\label{eq:FunctionZeta}
\end{equation}
Finally, this expression is used in \eqref{eq:Functionf}, and we obtain
\begin{align}
\label{eq:PhaseshiftContinuous}
f(+\infty) &= \\
&\hspace{-20pt}   \f{\omega + \f{i}{\hbar v_m} \int \D x \hspace{1mm} \lambda( x ) e^{i k x} \int_{x} \D x' \lambda(x') e^{- i k x'} }{\omega - \f{i}{\hbar v_m} \int \D x \hspace{1mm} \lambda( x ) e^{i k x} \int^{x} \D x' \lambda(x') e^{- i k x'}    } f(-\infty) \nonumber.
\end{align}
Comparing this to Eq.\@ \eqref{eq:FrequencyShift} proves the statement in Eq.\@ \eqref{eq:CouplingReplacement}.

\subsection{Multiple Majoranas Coupled to Each Edge}
\label{sec:CouplingMultipleVortices}
In the main text, we obtained a phase contribution of $\phi = \arctan(\f{\nu}{eV})$ in the differential conductance when the edge was coupled to one vortex (Eq.\@ \eqref{eq:DifferentialConductanceOne}). This can be generalized if the chiral Majoranas have single-point couplings to several vortices. In that case $\phi$ is replaced by $\sum_i \phi_i - \sum_j \phi_j$, where $\phi_i = \arctan(\f{\nu_i}{eV})$ comes from vortex-edge-coupling with the upper edge and $\phi_j$ from coupling with the lower edge. Hence, multiple phase crossovers will occur if several vortices are located close to the edge.

\subsection{One Majorana Coupled to Both Edges}
\label{sec:CouplingBothEdges}
Another generalization is to study point-couplings to both the lower and the upper arm, in which case the coupling term in the Lagrangian is $ \pazocal{L}_{\mathrm{bulk}-\mathrm{edge}} = 2 i \lambda_1 \xi_1(x=0) \xi_0 + 2 i \lambda_2 \xi_2(x=0)\xi_0$. The corresponding equations of motion are
\begin{align}
2\partial_t \xi_0 &= \lambda_1\left[ \xi_{1R} + \xi_{1L} \right]  + \lambda_2\left[ \xi_{2R} + \xi_{2L} \right], \\
v_m \xi_{1R} &= v_m \xi_{1L} + \lambda_1 \xi_0, \\
v_m \xi_{2R} &= v_m \xi_{2L} + \lambda_2 \xi_0.
\label{eq:EquationsOfMotion}
\end{align}
Here, we again use the notation $\xi_{1R} = \xi_1(x=0^+)$, $\xi_{2R} = \xi_2(x=0^+)$, $\xi_{1L} = \xi_1(x=0^-)$, and $\xi_{2L} = \xi_2(x=0^-)$. In frequency space we find a relation between $\xi_1$ and $\xi_2$ across the coupling points, $[\xi_{1}, \xi_{2}]_R^T = U(\omega)[\xi_{1}, \xi_{2}]_L^T$, where $U(\omega)$ is the unitary matrix
\begin{equation}
\label{eq:UnitaryMat}
U(\omega) = 
\f{1}{W(\omega)}\begin{pmatrix}
-\nu_1 + \nu_2 + i\omega & -2\sqrt{\nu_1\nu_2} \\
-2\sqrt{\nu_1\nu_2} & \nu_1 - \nu_2 + i\omega
\end{pmatrix}.
\end{equation}
Above, $\nu_i \equiv \lambda_i^2/(2\hbar v_m)$ and $W(\omega) = \nu_1 + \nu_2 + i\omega$. Notice how $U(\omega)$ is off-diagonal in the low-energy limit for the configuration $\nu_1 = \nu_2$. The chiral Majoranas on the perimeter therefore switch place by tunnelling across the vortex Majorana in this case. The phase matrix is given by
\begin{equation}
\pazocal{P} =
\begin{pmatrix}
e^{i\f{kl_1}{2}} & 0 \\
0 & e^{i \f{k l_2}{2}}
\end{pmatrix}
U(\omega)
\begin{pmatrix}
e^{i \f{k l_1}{2}+ in \pi} & 0 \\
0 & e^{i\f{k l_2}{2}}
\end{pmatrix},
\label{eq:PhaseMatrixTwoCoupling}
\end{equation}
which leads to the differential conductance
\begin{align} \label{eq:DifferentialConductanceTwo}
G_{\mathrm{SC}}(V) &= \f{2e^2}{h} \f{1}{(eV)^2 + (\nu_1 + \nu_2)^2} \Big[ (\nu_1 - \nu_2)^2 \nonumber \\
& \hspace{10pt} + \left[(eV)^2 - (\nu_1 - \nu_2)^2 \right] \sin^2\Big( \f{n\pi}{2}  + \f{eV \delta L}{2\hbar v_m} \Big) \nonumber \\
& \hspace{10pt} + eV(\nu_1-\nu_2)\sin\Big( n\pi + \f{eV \delta L}{\hbar v_m} \Big) \\
& \hspace{10pt} + 2\nu_1\nu_2(1+(-1)^n) \Big] \nonumber.
\end{align}
This reduces to Eq.\@ \eqref{eq:DifferentialConductanceOne} when $\nu_2 = 0$. For $\delta L = 0$ the expression above is identical to Eq.\@ \eqref{eq:DifferentialConductanceOne} with the replacement $\nu \to \nu_1 + \nu_2$.

\subsection{One Majorana Coupled to One Edge at Finite Temperature}
\label{sec:CouplingFiniteTemperature}
We return to the case with a one-point Majorana coupling (Fig.\@ \ref{fig:VortexEdgeCouplingSchematic} (a)). The drain current can generally be re-expressed as a residue sum over poles $z_j^{+}$ in the upper complex half-plane. Rewriting Eq.\@ \eqref{eq:DrainCurrent}, 
\begin{equation}
I_D(\nu) = (-1)^n \f{e}{4}i\sinh(\beta eV) \sum_{j} \mathrm{Res
}\lbrace h(z), z_j^{+} \rbrace,
\label{eq:ResidueSum}
\end{equation}
where $h(z)$ is the function
\begin{equation}
h(z) = \f{1}{z^2+\nu^2} \f{(z^2-\nu^2)\cos(\f{\delta L z}{v_m}) -2z\nu \sin(\f{\delta L z}{v_m})}{\cosh[\f{\beta}{2}(z-eV)] \cosh[\f{\beta}{2}(z+eV)]}.
\label{eq:ResidueSumFunction}
\end{equation}
The set of simple poles of $h(z)$ in the upper half-plane is $z_j^{+} \in \lbrace i\nu \rbrace \cup \lbrace i\pi(2m-1)/\beta + eV \rbrace_{m=1}^{\infty}$. For $\nu = 0$, the sum is obtainable in closed form and stated in Eq.\@ \eqref{eq:CurrentZerothOrder}.

For weak coupling at finite temperature, $\nu \ll k_B T, eV$, we expand $h(z)$ in powers of $\beta \nu$, with $\beta^{-1} = k_B T$. For convenience we define the reduced current as $R \equiv I_D(\nu)/I_{D,0}$. To second order in $\beta \nu$ with $\delta L = 0$ we find
\begin{align}
R &= 1 - \f{\nu \pi}{eV} \tanh\left(  \beta eV/2 \right) \label{eq:ReducedCurrentSecondOrder} \\
&  + \f{ \beta \nu^2 }{\pi eV} \Im\Big\lbrace \psi^{(1)}\Big(\f{1}{2} - i \f{\beta \nu}{2\pi} \Big) \Big\rbrace + \pazocal{O}\left( \beta \nu \right)^3. \nonumber
\end{align}

Above, $\psi^{(1)}(z) = \f{\D^2}{\D z^2} \log \Gamma(z)$ is the trigamma function. In the infinite coupling limit the sign of the drain current is flipped, $R \to -1$, which can be seen from Eq.\@ \eqref{eq:ResidueSumFunction}. The reduced current decreases monotonically, a feature of setting $\delta L = 0$, with coupling strength until the vortex Majorana is fully absorbed by the edge. 
\begin{figure}[h!tb]
\centering
\includegraphics[width=0.70\linewidth]{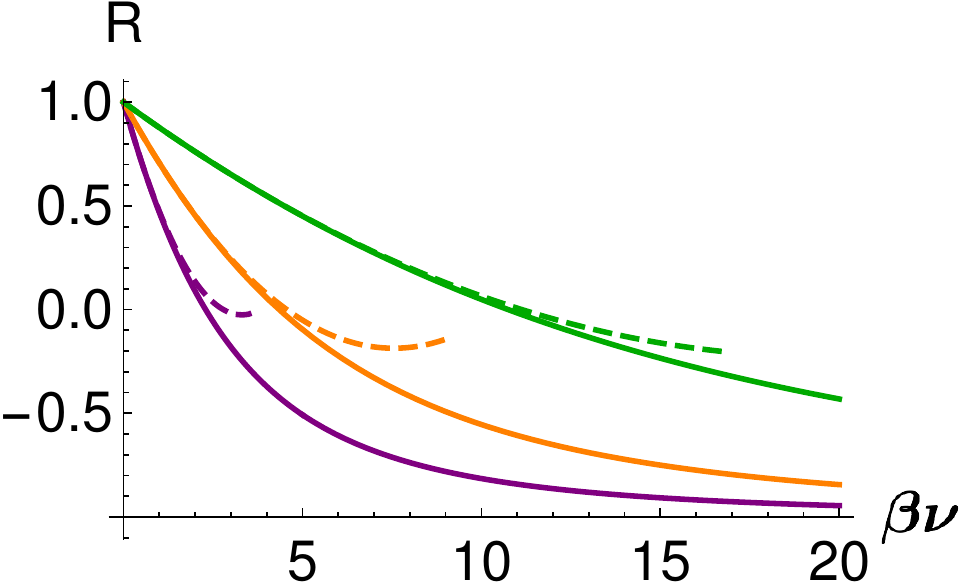}
\caption{The reduced drain current $R$ as function of $\beta \nu$ for symmetric arms $\delta L = 0$. The dotted lines represent the weak coupling result from Eq.\@ \eqref{eq:ReducedCurrentSecondOrder} and the full lines are numerical results for $\beta eV$ being $5$ (purple), $10$ (orange), and $25$ (green).}
\label{fig:RedCurrent}
\end{figure}
At $T = 0$ the current is found to be (with $\hbar = 1$ here):
\begin{align}
 (-1)^n\f{2\pi}{e} I_D &=  \f{v_m}{\delta L} \sin\Big( \f{\delta L eV}{v_m}\Big) \nonumber \\
 &\hspace{-20pt} + 2\nu \Big[ \cosh\Big( \f{\delta L eV}{v_m} \Big) - \sinh\Big( \f{\delta L eV}{v_m} \Big) \Big] \\ 
   &\hspace{-50pt} \times \Big[ \Im \Big\lbrace \mathrm{Ci}\Big( \f{\delta L}{v_m}(eV + i \nu) \Big) \Big\rbrace  + \Im \Big\lbrace \mathrm{Ci}\Big( -i \nu \f{\delta L}{v_m}\Big) \Big\rbrace \nonumber \\
 &\hspace{-20pt} - \Re \Big\lbrace \mathrm{Si}\Big( \f{\delta L}{v_m}(eV + i \nu) \Big) \Big\rbrace \Big] \nonumber.
\label{eq:CurrentatzeroTemp}
\end{align}
Here, $\mathrm{Ci}$ and $\mathrm{Si}$ are trigonometric integral functions. For symmetric arms the above result simplifies to give the reduced current $R = 1+ 2\f{\nu}{eV}[\arctan{(\f{\nu}{eV})} -\f{\pi}{2}]$. From this, the aforementioned result $\lim_{\nu \to \infty} R = -1$ follows trivially.


\section{Majorana Fermions in a TI/SC Hybrid Structure}
\label{sec:MajoranaBoundStates}
Superconductivity induced on the surface of a strong TI support chiral Majorana fermions on domain walls and Majorana bound states localized in vortices. The Fu and Kane Hamiltonian of a TI/SC hybrid structure with a single Dirac-like dispersion is given by \cite{FuKane_proximity, FuKane_interferometer}
\begin{equation}
\pazocal{H} = (v_F \boldsymbol{\sigma} \cdot \boldsymbol{p}-\mu) \tau_z + M(\bo{r}) \sigma_z  +  \Delta(\bo{r}) \tau_{+} + \Delta^{\ast}(\bo{r}) \tau_{-},
\label{eq:Model_Hamiltonian}
\end{equation}
where $\tau_j$ and $\sigma_j$ are Pauli matrices acting in particle-hole and spin space, respectively. Moreover, $\tau_{\pm} = (\tau_{x} \pm i \tau_y)/2$ and $M$ is the Zeeman energy associated with a magnetic field applied in the $z$ direction. The standard procedure is to introduce $u_{\sigma}(\boldsymbol{r})$ and $v_{\sigma}(\boldsymbol{r})$ that define the quasiparticles, conveniently arranged in  $\boldsymbol{\Psi} = [u_{\uparrow}, u_{\downarrow}, v_{\uparrow}, -v_{\downarrow}]^T$ satisfying the BdG equations $\pazocal{H} \boldsymbol{\Psi} = E \boldsymbol{\Psi}$.

\subsection{Corrections to the $S$ Matrix}
\label{sec:SmatrixCorrections}
One may solve Eq.\@ \eqref{eq:Model_Hamiltonian} on a magnetic domain wall in the absence of a superconductor. There are then \emph{two} chiral solutions: one in the particle sector, $\ket{\tau_z = +1}$, and one in the hole sector, $\ket{\tau_z = -1}$, both localized at the interface.\cite{FuKane_interferometer} On a magnet-superconductor interface there exists a single chiral Majorana solution with linear dispersion. By the definition of the $S$ matrix, 
\begin{equation}
\begin{pmatrix}
\xi_1 \\ \xi_2 
\end{pmatrix} = \begin{pmatrix}
S_{ee} & S_{eh} \\
S_{he} & S_{hh}
\end{pmatrix} \begin{pmatrix}
\psi_e \\ \psi_h
\end{pmatrix},
\label{eq:SmatrixDef}
\end{equation}
one can estimate the scattering elements at the trijunction as overlaps between the chiral states, $S_{ee} = \braket{\tau_z = +1}{\xi_1}$, $S_{eh} = \braket{\tau_z = -1}{\xi_1}$, etc. Here, $\xi_1$ ($\xi_2$) is the chiral Majorana fermion on the upper (lower) interferometer arm. The Majorana wavefunction decays exponentially with the length scale $v_F/\sqrt{M^2-\mu^2}$ ($v_F/\Delta_0$) in the magnetic (superconducting) region (inset of Fig.\@ \ref{fig:WavefuncsAndRate} (a)). Interestingly, if the magnetization on each side of the Dirac channel are of equal magnitude (with opposite polarization), compactly expressed in terms of the symmetry $\pazocal{H}(-y) = \pazocal{M}^{-1}\pazocal{H}(y) \pazocal{M}$ where $\pazocal{M} = i\sigma_y$,\cite{FuKane_interferometer} the zero energy result in Eq.\@ \eqref{eq:SzeroEnergy} is obtained at all energies. Experimentally, this is likely a weak assumption, and we can therefore safely apply $S(E=0)$ throughout. If the two magnetic fields in the Dirac region are of different strengths there will be corrections to $S$ that are small when $E \ll \f{v_m}{v_F}\min{\lbrace M_0, \Delta_0 \rbrace}$. One can heuristically add a correction term to $S(E)$ proportional to $E/\Lambda$ (where $\Lambda \sim \f{v_m}{v_F}\min{\lbrace M_0, \Delta_0 \rbrace}$ is some phenomenological scale) and impose unitarity to order $\pazocal{O}(E^2)$ and particle-hole symmetry. This leads to order $\pazocal{O}(E^2)$ corrections to the drain current.

\subsection{Majorana Bound States and Energy Splitting}
\label{sec:VortexBoundStates}
If a vortex is present in the system described by Eq.\@ \eqref{eq:Model_Hamiltonian}, $\Delta(\bo{r}) = \Delta(r) e^{i\ell \theta}$, a single Majorana bound state will be localized in the vortex core when the vorticity $\ell$ is odd.\cite{TunnelingMajoranamodes, MajoranaSupercondIslands} Following the procedure in Ref.\@ \onlinecite{Sau_generalapproach}, the zero energy BdG equations for the model in Eq.\@ \eqref{eq:Model_Hamiltonian} with a central vortex of vorticity $\ell = 1$ are expressed in terms of two real and coupled radial equations,
\begin{equation}
\begin{pmatrix}
- \mu & \eta \Delta(r) + v_F\left( \partial_r + \f{1}{r} \right) \\
-\eta \Delta(r) - v_F\partial_r  &  - \mu
\end{pmatrix}
\begin{pmatrix}
u_{\uparrow}(r) \\
u_{\downarrow}(r)
\end{pmatrix}  = 0.
\label{eq:BdGZeroRadial}
\end{equation}
Here, $v_{\sigma}(r) = \eta u_{\sigma}(r)$ with $\eta = \pm 1$ dictating two possible solution channels. Modelling the vortex by a hard step of size $R_1 \approx \xi$, $\Delta(r) = \Delta_0 \Theta(r - R_1)$, the resulting zero mode is expressed in terms of Bessel functions,
\begin{align}
\begin{pmatrix}
u_{\uparrow}(r), 
u_{\downarrow}(r)
\end{pmatrix}^T &= \\
&\hspace{-60pt} \pazocal{N} \begin{pmatrix}
J_0(\f{\mu r}{v_F}) \\ J_1(\f{\mu r}{v_F})
\end{pmatrix} \left[ \Theta(R_1-r) + e^{-\f{\Delta_0}{v_F}(r-R_1)} \Theta(r-R_1)  \right] \nonumber.
\label{eq:SolutionVortexCore}
\end{align}
Above, $\pazocal{N}$ is given by normalization of the two-component spinor. Two vortex bound Majoranas separated by a distance $R$ will generally lift from zero energy by an energy splitting exponentially small in $R/\xi$ when $R \gg \xi$. If $\mu$ is tuned close to the Dirac point (typically achieved by bulk doping of screened charges\cite{BandFluctuationsPuddlesNat}), this splitting has previously been estimated to asymptotically approach $\varepsilon_+ \sim - \mu (R/\xi)^{3/2} \exp(-R/\xi)$ up to some prefactor of order one.\cite{TunnelingMajoranamodes} 


Similarly, a superconducting disk of radius $R$ (with a central vortex) deposited on a TI supports a Majorana edge state located exponentially close to the edge. For weak magnetic fields the splitting between the central bound state and the edge state has been estimated to decay like \cite{MajoranaSupercondIslands}
\begin{equation}
\varepsilon_{+} \sim - \mu l_B e^{-\f{R}{\xi} }  / \xi,
\label{eq:EnergySplit}
\end{equation}
with $l_B$ the magnetic length.

\subsection{Toy Model Calculation: Energy Splitting in a Spinless $p$-wave Superconductor}
\label{sec:Energysplittingpplusip}
For completeness, and serving as an illuminating example not found elsewhere, we explain how the energy splitting between edge states in a Corbino geometry can be calculated in the spinless $p+ip$ superconductor. See e.g.\@ Ref.\@ \onlinecite{fermMajo12} for general aspects of this system and Ref.\@ \onlinecite{PRLMajoranaSplitting} for a presentation of the similar intervortex splitting. Finally, the problem of the Majorana energy hybridization in superconductor-semiconductor hybrid structures is addressed in Ref.\@ \onlinecite{MajoranaQubitDephasing}.

The spinless $p$-wave superconductor realises a topological phase for $\mu > 0$ and the trivial phase for $\mu < 0$. The model has the corresponding zero energy BdG equations\cite{AnyonQuantumComputation}
\begin{equation}
\begin{pmatrix} -\f{1}{2m}\nabla^2 - \mu & \f{1}{2 p_F}\{\Delta(\bo{r}),\partial_{z^{\ast}} \} \\ -\f{1}{2 p_F}\{\Delta^{\ast}(\bo{r}),\partial_{z} \} & \f{1}{2m}\nabla^2 + \mu \end{pmatrix}\begin{pmatrix} u(\bo{r}) \\ v(\bo{r}) \end{pmatrix} = 0,
\label{eq:BdGSpinlessp+ip}
\end{equation}
with the operator $\partial_{z^{\ast}} = e^{i\theta}(\partial_r + \f{i}{r}\partial_{\theta})$. We let a radial annulus geometry between $R_1$ and $R_2$ be superconducting with a vortex located in the center hole, $\Delta(\bo{r}) = \Delta_0 e^{i\theta} \Theta(r-R_1) \Theta(R_2-r)$, with chemical potential $2m v_F^2 \mu > \Delta_0^2$ (causing the wavefunctions to oscillate) and $\mu < 0$ outside the disk. When $R \equiv R_2 - R_1 \gg \xi$, we treat the two single-edged systems separately and construct the ground state candidates $\bo{\phi}_{\pm} = \f{1}{\sqrt{2}}(\bo{\psi}_1\pm i\bo{\psi}_2)$, where $\bo{\psi}_1$ ($\bo{\psi}_2$) is the solution localized on the inner (outer) edge.\cite{PRLMajoranaSplitting} The two solutions are exponentially damped away from their respective edges and they oscillate with frequency $k = \sqrt{2m\mu - (\Delta_0/v_F)^2}$. Moreover, if both $R_1$, $R_2 \gg 1/k$, the Bessel functions that appear as radial solutions can be expanded asymptotically. The splitting integral can be calculated, and we send $\mu \to -\infty$ outside the annulus in the end. Upon evaluating the splitting integrals and then taking the limit, we obtain 
\begin{equation}
\varepsilon_{\pm} = \expval{\pazocal{H}}{\bo{\phi}_{\pm}} \approx \mp \f{4\Delta_0 \mu}{v_F k} \sin \left( k R \right)  e^{- R/\xi }.
\label{eq:Corbino_E+}
\end{equation}
The splitting is zero for particular values of the domain separation. Whenever $R = \pi n /k$ for $n \in \mathbb{N}$, there are two degenerate ground states. The zero mode condition gives associations of an interference phenomenon, caused by the oscillating edge modes that convolute in a destructive manner for certain values of the separation. The expression in Eq.\@ \eqref{eq:Corbino_E+} agrees with numerical diagonalization, already when $R$ is a small multiple of $\xi$. The Corbino geometry has been studied for small disk size, in which case the same zero energy criterion is found in the limit $2m v_F^2 \mu \gg \Delta_0^2$ by imposing Dirichlet boundary conditions on the wavefunctions directly.\cite{Majoranamodes_disk}

\section{Surface-Bulk Scattering with Screened Disorder}
\label{sec:SBScatteringAppendix}
In Ref.\@ \onlinecite{SBcouopling14} Fermi's Golden rule is used to find the scattering rate from the surface ($S$) to the bulk ($B$) of a TI in the presence of static and dilute point impurities. Here, we take Ref.\@ \onlinecite{SBcouopling14} as a starting point (the reader is referred to this reference for further details) and apply the formalism to the case of long range scatterers. As described in the main text we study scattering from surface initial wave vector $\bo{k} = (\bo{k}_{\parallel}, 0)$ (in practice we let $\bo{k}_{\parallel} = (k_x, 0)$) to the final bulk wave vector $\bo{k}' = (\bo{k}_{\parallel}', k_z')$. The energy of the incoming surface state determines the Fermi energy, $\xi_{S,\bo{k}_{\parallel} } = v_F  k_{\parallel} \equiv \epsilon_F$. Assuming low energy elastic scattering, and making use of the continuum limit $\sum_{\bo{k}'} \to \f{V}{(2\pi)^3} \int \D^3\bo{k}'$, we find the scattering rate
\begin{widetext}
\begin{align}
\Gamma^{\mathrm{imp}}_{\mathrm{SB}}(\epsilon_F) &= 2\pi \sum_{\bo{k}', n'} \lvert g^{\mathrm{imp}}_{\bo{k}-\bo{k}'} \rvert^2   \big(  \lvert F_{S,\bo{k}_{\parallel}; B, \bo{k}',1n' } \rvert^2  + \lvert F_{S,\bo{k}_{\parallel}; B, \bo{k}', 2n' } \rvert^2   \big)    \delta(  \xi_{B,n', k_{\parallel}' } - \epsilon_F  ) \label{eq:ScatteringRateFull}  \\
&\approx \frac{ V }{(2\pi)^2} \sum_{\substack{n' : \\ \min \lbrace \xi_{B,n', k_{\parallel}' } \rbrace < \epsilon_F  } }  \f{ k_{\parallel}' }{\lvert \nabla  \xi_{B,n', k_{\parallel}' }  \rvert}  \int_0^{\infty} \D k_z'  \int_0^{2\pi} \D \varphi \hspace{1mm} \lvert g^{\mathrm{imp}}_{ k_{\parallel} - k_{\parallel}', k_z', \varphi} \rvert^2   \big(  \lvert F_{S,\bo{k}_{\parallel}; B, \bo{k}',1n' } \rvert^2  + \lvert F_{S,\bo{k}_{\parallel}; B, \bo{k}', 2n' } \rvert^2   \big). \nonumber
\end{align}
\end{widetext}
In going from the first to the second line we integrated over $k_{\parallel}'$, which in the last line is then defined implicitly by $\xi_{B,n', k_{\parallel}' } = \epsilon_F$. Above, $\varphi$ is the polar angle between the incoming and the outgoing momentum projected onto the $k_x k_y$-plane, i.e. $k_{x}' = k_{\parallel}' \cos{\varphi}$ and $k_{y}' = k_{\parallel}' \sin{\varphi}$. The $F$'s are defined as convoluted overlaps between the surface and the bulk states, 
\begin{align}
F_{S,\bo{k}_{\parallel}; B, \bo{k}',1n' } &= \big<\bo{\Psi}_{S, \bo{k}_{ \parallel}} \big|   e^{ik_z' z} \big| \bo{\Psi}_{B, \bo{k}', 1n'} \big>,  \label{eq:Overlap1} \\
F_{S,\bo{k}_{\parallel}; B, \bo{k}',2n' } &= \big<\bo{\Psi}_{S, \bo{k}_{ \parallel}} \big|   e^{ik_z' z} \big| \bo{\Psi}_{B, \bo{k}', 2n'} \big>, \label{eq:Overlap2}
\end{align}
where two bulk bands $1n'$ and $2n'$ are degenerate. The four-component wavefunctions in these expression are found by solving the BdG equations for the 3D TI exactly at $k_{\parallel} = 0$ and then applying perturbation theory to leading order in the wave vector (see Appendix C in Ref.\@ \onlinecite{SBcouopling14} where the full expressions are listed in Eq.\@ (C1)--(C11)). This procedure also yields the dispersion relations to leading order in the parallel wave vector. In Fig.\@ \ref{fig:WavefuncsAndRate} (a) and (b) the surface and some of the bulk states are visualised in a thin film with model parameters as for Bi$_2$Se$_3$,\cite{HamiltonianSurface10} except for the bulk gap which is set to $\Delta_{\mathrm{TI}} = 0.1$ eV as motivated in the main text.

The coupling $g^{\mathrm{imp}}_{\bo{k}-\bo{k}'}$ is here an ensemble averaged Coulomb potential due to screened charge impurities. Assuming randomly distributed impurities with zero mean, the coupling should be proportional to $\lvert g^{\mathrm{imp}}_{\bo{k}-\bo{k}'} \rvert^2 \propto n_{\mathrm{3D}} ( Ze^2/(\epsilon_0 \epsilon_r) )^2 \left( \lvert \bo{k}-\bo{k}' \rvert^2 + k_{TF}^2 \right)^{-2}$,\cite{ElectronsAndPhonons} where $\epsilon_r$ is the relative permittivity, $k_{TF}$ is the Thomas-Fermi wave vector, and $n_{\mathrm{3D}}$ is the average dopant concentration. In cylindrical coordinates the coupling is expressed as
\begin{align}
\label{eq:ScatteringPotentialScreened}
\lvert g^{\mathrm{imp}}_{k_{\parallel} - k_{\parallel}', k_z',\varphi} \rvert^2 &= \f{n_{\mathrm{3D}}}{V} \left( \f{Ze^2}{\epsilon_0 \epsilon_r k_{\parallel}^2} \right)^2 \\
& \hspace{-50pt} \times  \left[ \left(1- k_{\parallel}' / k_{\parallel} \right)^2 + \left( k_z' / k_{\parallel} \right)^2 + r_s^2 + 4 \f{k_{\parallel}'}{k_{\parallel}} \sin^2{\f{\varphi}{2}}   \right]^{-2}, \nonumber
\end{align} 
where $r_s = k_{TF}/k_{\parallel}$ was introduced. In the main text we assume that the screened charges have $Z = 1$. With this coupling established we define the differential cross section for the screened Coulomb scattering as
\begin{align}
\f{\D \sigma_{\mathrm{long}}^{(n')}(\epsilon_F) }{\D k_z'} &\equiv \label{eq:DifferentialCrossectionLong}  \\
& \hspace{-40pt} \int_0^{2\pi} \D\varphi \hspace{1mm} \f{  \lvert F_{S,\bo{k}_{\parallel}; B, \bo{k}',1n' } \rvert^2  + \lvert F_{S,\bo{k}_{\parallel}; B, \bo{k}', 2n' } \rvert^2   }{\left[ \left(1- \f{k_{\parallel}'}{ k_{\parallel} } \right)^2 + \left( \f{k_z'}{ k_{\parallel} } \right)^2 + r_s^2 + 4 \f{k_{\parallel}'}{k_{\parallel}} \sin^2{\f{\varphi}{2}}   \right]^{2}} \nonumber.
\end{align}
This function is shown in Fig.\@ \ref{fig:WavefuncsAndRate} (c) for $r_s = 0.035$, which corresponds to $\epsilon_r = 100$.\cite{SurfImpScat10}  Note that in the case of point scatterers, the differential cross section is obtained simply by replacing the denominator in Eq.\@ \eqref{eq:DifferentialCrossectionLong} by $1$.

Using Eq.\@ \eqref{eq:ScatteringPotentialScreened} and \eqref{eq:DifferentialCrossectionLong} in \eqref{eq:ScatteringRateFull} assembles the expression for the scattering rate in the main text, Eq.\@ \eqref{eq:ScatteringRateLong}.   We note that the rates as shown in figure in Fig.\@ \ref{fig:WavefuncsAndRate} (d) are seen to be in good agreement with the Ref.\@ \onlinecite{SBcouopling14}.

\subsection{Effect of Spatially Confined Transverse Wavefunction}
\label{sec:ConfinedWF}
When applying the formula of the electronic scattering rate in Eq.\@ \eqref{eq:ScatteringRateLong} to the case of the surface Majorana we neglect the transverse profile of the surface wavefunction (inset of Fig.\@ \ref{fig:WavefuncsAndRate} (a)). Here, we argue why this is a valid approximation in our parameter regime. 

Assume for simplicity that the two surface gaps are of similar magnitude $\Delta_0 \approx M_0 \gg \mu$, so that the the transverse wavefunction is confined over a the scale $\sigma_y \sim \xi = v_F/\Delta_0$.  By the uncertainty principle, this confinement leads to an uncertainty in $k_y'$ of $\sigma_{k_y'} \sim \xi^{-1}$. To simulate this uncertainty we draw random additions to $k_y' = k_{\parallel}' \sin \varphi$ from a normal distribution with the standard deviation above and zero mean for each scattering direction $\varphi$. The resulting scattering rates display an average absolute deviation $\langle \lvert \varepsilon \rvert \rangle$ (averaged over the Fermi energy range in Fig.\@ \ref{fig:WavefuncsAndRate} (d)) from the sharp $k_y'$ curve of roughly $\langle \lvert \varepsilon \rvert \rangle \approx (\sigma_{k_y'}/\mathrm{max}\lbrace k_y' \rbrace )^2$. Since $ \mathrm{max}\lbrace k_y' \rbrace \gg \xi^{-1}$ in the energy range and with the proximity gaps we consider, the effect of uncertainty in $k_y'$, and hence confinement in the $y$-direction, can be ignored.

\section{Scattering with Acoustic Phonons}
\label{sec:AcousticPhonons}
Consider the toy model coupling electrons to (acoustic) phonons through a deformation potential,
\begin{equation}
H_{\mathrm{ep}} = \sum_{\bo{q}_1,\bo{q}_2,s, s'} M_{\bo{q}_1,\bo{q}_2}^{s,s'} c_{\bo{q}_1+\bo{q}_2,s}^{\dagger} c_{\bo{q}_1,s'}\left( a_{\bo{q}_2} + a_{-\bo{q}_2}^{\dagger} \right).
\label{eq:ElectronPhononCoupling}
\end{equation}
For small momenta, the electron-phonon coupling goes as $\lvert M(q) \rvert \sim q$ for surface phonons.\cite{ElectronPhononScatteringTI, PhononScatteringvonOppen} Acoustic phonons have linear dispersion, $\omega(q) = c_R \lvert \bo{q} \rvert$ and are expected to dominate the coupling Hamiltonian at low temperatures.\cite{ElectronPhononScatteringTI} The scattering rate follows once again from Fermi's Golden rule, $\Gamma_{i\to f}^{\mathrm{ph}} = 2\pi \nu_f(E) \lvert \mel{f}{H_{\mathrm{ep}}}{i} \rvert^2$, where $\nu_f(E)$ is the final density of states. Recall also that states with a linear dispersion in two dimensions have $\nu(E) \propto E$. In a superconducting system, there will be a non-zero amplitude for the creation of a phonon with the cost of annihilating two quasiparticles. For illustrative purposes, we consider quasiparticles excitations of the ($s$-wave) Bardeen-Cooper-Schrieffer (BCS) ground state $\ket{\Omega}$, 
\begin{align}
\ket{i;\sigma_1, \sigma_2} &= \gamma_{\bo{k}_1,\sigma_1}^{\dagger} \gamma_{\bo{k}_2,\sigma_2}^{\dagger} \ket{\Omega}, \label{eq:InitialState} \\
\ket{f} &= a_{\bo{q}}^{\dagger} \ket{\Omega},  \label{eq:FinalState} \\
\ket{\Omega} &= \prod_{\bo{k}} (u_{\bo{k}} + v_{\bo{k}}c_{\bo{k},\uparrow}^{\dagger}c_{-\bo{k},\downarrow}^{\dagger})\ket{0}.
\label{eq:BCSGroundstate}
\end{align}
Above, the quasiparticle creation operators are $\gamma_{\bo{k},+}^{\dagger} = u_{\bo{k}}^{\ast} c_{\bo{k},\uparrow}^{\dagger} - v_{\bo{k}}^{\ast} c_{-\bo{k},\downarrow}$ and $\gamma_{\bo{k},-}^{\dagger} = u_{\bo{k}}^{\ast} c_{-\bo{k},\downarrow}^{\dagger} + v_{\bo{k}}^{\ast} c_{\bo{k},\uparrow}$.
The scattering element is found to be
\begin{widetext}
\begin{equation}
\begin{aligned}
\mel{f}{H_{\mathrm{ep}}}{i;\sigma_1, \sigma_2} &=  \Big( M_{\sigma_1 \bo{k}_1, -\sigma_1\bo{k}_1 - \bo{k}_2}^{\uparrow, s(\sigma_1)} \delta_{\bo{q},\sigma_1 \bo{k}_1 + \bo{k}_2} - M_{\sigma_1 \bo{k}_1,-\sigma_1\bo{k}_1 + \bo{k}_2}^{\downarrow, s(\sigma_1)} \delta_{\bo{q},\sigma_1 \bo{k}_1 - \bo{k}_2}  \Big) v_{-\bo{k}_2}^{\ast} u_{\bo{k}_1}^{\ast} \\
&\hspace{10pt} - \Big( M_{\sigma_2 \bo{k}_2, -\sigma_2\bo{k}_2 - \bo{k}_1}^{\uparrow, s(\sigma_2)} \delta_{\bo{q},\sigma_2 \bo{k}_2 + \bo{k}_1}  - M_{\sigma_2 \bo{k}_2,-\sigma_2\bo{k}_2 + \bo{k}_1}^{\downarrow, s(\sigma_2)} \delta_{\bo{q},\sigma_2 \bo{k}_2 - \bo{k}_1} \Big) v_{-\bo{k}_1}^{\ast} u_{\bo{k}_2}^{\ast}.
\end{aligned}
\label{eq:Scattering_Majorana_withSpin}
\end{equation}
\end{widetext}
Above, we introduced the symbol $s(+) = \hspace{1mm} \uparrow$ and $s(-) = \hspace{1mm}  \downarrow$. The scattering amplitude depends only on the momentum transfer for small energies, in which case the coupling above vanishes identically. This is presumably because the average charge of the quasiparticles becomes zero in the low energy limit. 

A careful analysis for the electron decay rate alone yields a $\Gamma^{\mathrm{ph}} \sim T^3$ law well below the Bloch-Gr\"{u}neisen temperature at the Fermi surface.\cite{ElectronPhononScatteringTI} Away from the Fermi surface, by biasing the electrons at a small voltage $V$, the decay rate is finite at $T= 0$ and goes as $\Gamma^{\mathrm{ph}} \sim (eV)^3$. Inserting the exact prefactor (for Bi$_2$Te$_3$), by following the steps in Ref.\@ \onlinecite{ElectronPhononScatteringTI}, we find a decay rate in the peV range when $V \simeq 1$ $\mu$V. This means that the electron lifetime due to acoustic phonon scattering $\tau_{\mathrm{ph}} = 1/\Gamma^{\mathrm{ph}}$ is in the ms range, making this effect negligible at $T = 0$ at low voltage. 

\end{appendix}

\bibliography{ReferencesMajorana}

\end{document}